\documentclass[a4paper,11pt]{article}
\pdfoutput = 1
\usepackage{jheppub}

\usepackage{amsmath,amssymb,amscd,braket,amsfonts}
\usepackage{color}
\usepackage{graphicx}
\usepackage{slashed}
\usepackage{amsthm}

\newcommand{\bea}{\begin{eqnarray}}
\newcommand{\eea}{\end{eqnarray}}
\newcommand{\be}{\begin{equation}}
\newcommand{\ee}{\end{equation}}
\newcommand{\ba}{\begin{align}}
\newcommand{\ea}{\end{align}}

\title{Convergence of the Fefferman-Graham expansion\\ and complex black hole anatomy}
\author[a]{Alexandre Serantes}
\author[b]{and Benjamin Withers}
\date{April 2022}

\affiliation[a]{Departament de Física Quàntica i Astrofísica, Institut de Ciències del Cosmos (ICCUB), Facultat de Física, Universitat de Barcelona, Martí i Franquès 1, Barcelona,  ES08028, Spain}
\affiliation[b]{Mathematical Sciences and STAG Research Centre, University of Southampton, Highfield, Southampton SO17 1BJ, UK}

\emailAdd{alexandre.serantes@ub.edu}
\emailAdd{b.s.withers@soton.ac.uk}

\abstract{
Given a set of sources and one-point function data for a Lorentzian holographic QFT, does the Fefferman-Graham expansion converge? If it does, what sets the radius of convergence, and how much of the interior of the spacetime can be reconstructed using this expansion?
As a step towards answering these questions we consider real analytic CFT data, where in the absence of logarithms, the radius is set by singularities of the complex metric reached by analytically continuing the Fefferman-Graham radial coordinate.
With the conformal boundary at the origin of the complex radial plane, real Lorentzian submanifolds appear as piecewise paths built from radial rays and arcs of circles centred on the origin. 
This allows singularities of Fefferman-Graham metric functions to be identified with gauge-invariant singularities of maximally extended black hole spacetimes, thereby clarifying the physical cause of the limited radius of convergence in such cases.
We find black holes with spacelike singularities can give a radius of convergence equal to the horizon radius, however for black holes with timelike singularities the radius is smaller.
We prove that a finite radius of convergence does not necessarily follow from the existence of an event horizon, a spacetime singularity, nor from caustics of the Fefferman-Graham gauge, by providing explicit examples of spacetimes with an infinite radius of convergence which contain such features.
}

\begin{document}
\maketitle
% \begin{titlepage}

% \begin{center}

% \end{center}

% \vfill

% \end{titlepage}

% \tableofcontents

\newpage
\section{Introduction}
The AdS/CFT correspondence is a celebrated duality between strongly coupled conformal field theories (CFTs) and asymptotically anti-de Sitter (AdS) gravity \cite{Maldacena:1997re,Gubser:1998bc,Witten:1998qj}. Given the one-point functions for a Lorentzian CFT in the presence of sources, what can be determined about the bulk geometry? Under the AdS/CFT dictionary and after appropriate renomalisation \cite{Henningson:1998ey,Henningson:1998gx, Haro2000,Skenderis:2000in,Bianchi2001,Skenderis:2002wp} the one-point functions and sources appear as near-boundary data. Attempting to reconstruct the bulk geometry by using the corresponding Cauchy problem with sources and vevs appearing as Cauchy data is not a well-posed problem in general, highlighting the main obstruction to answering this question. We investigate the bulk reconstruction question by asking,
\begin{quote}
\emph{Is the Fefferman-Graham near-boundary expansion convergent for smooth Lorentzian near-boundary data (sources and vevs)? Where convergent, what sets the radius of convergence -- a physical obstruction, or an artefact of Fefferman-Graham gauge? If factorially divergent, what non-perturbative contributions complete it?}
\end{quote}
In Fefferman-Graham gauge, asymptotically locally AdS$_{d+1}$ spacetimes take the form\footnote{We set the AdS radius $L=1$ throughout.}
\be
G = G_{AB} dX^A dX^B = \frac{du^2}{u^2} + \gamma_{\mu\nu}(u,x)dx^\mu dx^\nu, \label{metric}
\ee
with bulk indices $A,B = 0, 1, \ldots, d$ and boundary indices $\mu,\nu = 1, 2, \ldots, d$. The metric $\gamma_{\mu\nu}$ has a double pole at $u=0$, so that $\lim_{u\to 0} u^2 \gamma_{\mu\nu}$ is finite. This corresponds to a conformal boundary of the spacetime, or at least a portion of it. In the Fefferman-Graham expansion~\cite{FG}, one develops the following formal small $u$ expansion of $\gamma_{\mu\nu}$,
\be
u^2\gamma_{\mu\nu}(u,x) = h_{\mu\nu}(x) + \ldots,\label{FG}
\ee
where the ellipses denote subleading terms which vanish as $u\to 0$. Here $h_{\mu\nu}$ is a representative boundary metric for its conformal class, namely those metrics which are related up to an overall smooth, positive factor.  In the case of vacuum Einstein gravity Fefferman and Graham \cite{FG} introduced an expansion of the form,
\bea
u^2\gamma_{\mu\nu}(u,x) = h_{\mu\nu}(x) + h^{(2)}_{\mu\nu}(x)u^2 + \ldots +  t_{\mu\nu}(x)u^d + h^{(d)}_{\mu\nu}(x) u^d \log{u}  + \ldots, \label{FGGR}
\eea
where all coefficients in the $u$ expansion can determined uniquely by solving the Einstein equations order-by-order in $u$, once the data $h_{\mu\nu}(x), t_{\mu\nu}(x)$ are specified. The logarithms in \eqref{FGGR} are only present if $d$ is even. The data $t_{\mu\nu}$ must satisfy trace and transversality constraints \cite{Haro2000}. In the context of AdS/CFT the data $h_{\mu\nu}(x), t_{\mu\nu}(x)$ determine the CFT metric and expectation value of the CFT stress-tensor obeying appropriate Ward identities, with a precise identification made through the procedure of holographic renormalisation in which boundary counterterms are constructed \cite{Henningson:1998ey,Henningson:1998gx, Haro2000,Skenderis:2000in,Bianchi2001,Skenderis:2002wp}.

For the case of vacuum Einstein gravity in particular, there are several existing results concerning the properties of the expansion. Given a real analytic boundary metric and real analytic stress tensor data, the expansion \eqref{FGGR} converges. This has been shown for odd $d$ with $t_{\mu\nu}=0$ in \cite{FG} and has been extended to $t_{\mu\nu}\neq 0$ for the case of $d=3$ in \cite{anderson1} and to all $d$ in \cite{KICHENASSAMY2004268}. A discussion of these and related results can be found in \cite{Anderson:2004yi,deBoer:2004yu,Anderson:2006ax}.
In addition, there are various special cases for which the expansion truncates and therefore has an infinite radius of convergence.
In $d>2$ \eqref{FGGR} truncates at order $u^4$ provided the bulk metric has a vanishing Weyl tensor \cite{Skenderis:1999nb}. In $d>2$ \eqref{FGGR} truncates at order $u^4$ if $h_{\mu\nu}$ is conformally Einstein and $t_{\mu\nu} = 0$ \cite{ambient}. In $d=2$ \eqref{FGGR} truncates at order $u^4$ yielding locally AdS$_3$ metrics. For real analytic $pp$-wave boundary metrics with $d$ even the \eqref{FGGR} truncates at order $u^{d-2}$ \cite{Leistner:2008wz}.

In the presence of matter, one may also expect that the expansion converges for real analytic data. This is straightforwardly confirmed in the following special case. Consider a CFT$_d$ with a $\Delta = (d+1)/2$ scalar operator $O_\phi$, with an AdS$_{d+1}$ dual and a corresponding $m^2 = (1-d^2)/4$ scalar field, $\phi$. Treating the scalar in a probe approximation around the planar AdS$_{d+1}$ vacuum, i.e.
\be
\gamma_{\mu\nu} = \frac{\eta_{\mu\nu}}{u^2}, \qquad \phi(u,x^\mu) = u^\frac{d-1}{2}\Psi(u,x^\mu),
\ee
then the equation of motion and boundary conditions for $\Psi$ become
\bea
&&\partial_u^2 \Psi = -\eta^{\mu\nu}\partial_\mu \partial_\nu \Psi, \label{Cauchy1}\\
&&\Psi(0,x^\mu) = s(x^\mu), \qquad \partial_u\Psi(0,x^\mu) = v(x^\mu), \label{Cauchy2}
\eea
with $s, v$ corresponding to the CFT source and vev for the operator $O_\phi$.
Then by the Cauchy–Kovalevskaya theorem, if both $s$ and $v$ are real analytic functions, then the corresponding Cauchy problem \eqref{Cauchy1}, \eqref{Cauchy2} has a unique analytic solution near the boundary $u=0$ and thus the Fefferman-Graham expansion converges in some neighbourhood of $u=0$.\footnote{This argument generalises straightforwardly to a fixed AdS-Schwarzschild background.} In such analytic cases, the question of interest then becomes what physical feature of the spacetime (if any) sets the radius of convergence.\footnote{It is clear that the \emph{value} of $u$ where convergence fails is itself not especially interesting, since this can be changed by changing the conformal class representative. For example, performing a homogeneous Weyl rescaling dilates the entire $u$-plane, and one would have to form an invariant instead.}

In this paper we take real analytic source and vev data $(h_{\mu\nu}(x), t_{\mu\nu}(x))$ for a given CFT metric and state. This is either data handed to us from the CFT, or computed holographically by imposing regularity in the interior, or by preparing states using real-time holography \cite{Skenderis:2008dh,Skenderis:2008dg}\footnote{See \cite{Skenderis:2009ju} for an example where the real-time formalism is necessary to compute the correct CFT vevs.}.
We consider cases for which the expansion of $u^2 \gamma_{\mu\nu}$ in \eqref{FG} takes the form of a power series in $u$.\footnote{Our analyses also apply to Puiseux series, since after removing an overall factor these series are also power series with an appropriate change of variables.} We do not consider cases with logarithms. Then it is a theorem that the radius of convergence of \eqref{FG} is governed by singularities of the functions $\gamma_{\mu\nu}(u,x)$ analytically continued to the complex $u$-plane as holomorphic functions. These are points where $\gamma_{\mu\nu}(u,x)$ fail to be holomorphic, such as poles, branch points or essential singularities. Thus, we are led to study the class of metrics \eqref{metric} with $u \in \mathbb{C}$ and $x^\mu \in \mathbb{R}$.

We stress that from first principles it is the singularities of the \emph{metric components} in Fefferman-Graham gauge in the complex $u$-plane that dictate the convergence. This is a priori divorced from the usual notions of spacetime singularities in general relativity. On one hand, it is possible that the radius is set by a mere coordinate singularity, though we have not encountered this in our examples. On the other hand, we can have the more intriguing scenario where the spacetime has a (gauge-invariant) curvature singularity but the metric components are everywhere regular, and thus the Fefferman-Graham expansion converges everywhere. This scenario we have encountered and one generic mechanism we have identified is a metric which degenerates while remaining analytic. We find two explicit examples of this. First for the black holes in section \ref{sec:axion:critical} near the black hole singularity the metric takes the form, $G= -d\tau^2 + dt^2 + \tau^2 d\vec{x}_{d-1}^2$ and thus is degenerate at $\tau=0$ where there is a divergent Ricci scalar. In this case the Fefferman-Graham expansion converges everywhere in the $u$-plane, including the location of this black hole singularity. Motivated by this observation we also analyse holographic RG flows in section \ref{sec:rgflow} and consider conditions placed on the scalar potential for the existence of an analytic degenerate point in the metric. In this second class of examples we engineer the potential to produce Fefferman-Graham expansions which truncate at arbitrarily large order.
\\\\
The layout of the paper is as follows. In section \ref{sec:sections} we analyse how real submanifolds of the complex $u$-plane can be constructed and give rise to physical, Lorentzian spacetime metrics. This gives physical meaning to portions of the complex $u$-plane beyond the traditionally studied case of $u>0$ on the real line. We then go on to analyse black holes in vacuum in section \ref{sec:vacuum}, black holes with matter in \ref{sec:matter}, and holographic RG flows in \ref{sec:rgflow}. We finish with a discussion and outlook in section \ref{sec:discussion}. Throughout this paper we restrict our analyses to $d\geq 2$.

\section{Causal structure in the complex Fefferman-Graham plane\label{sec:sections}}
In the absence of logarithms, singularities of $\gamma_{\mu\nu}$ in the complex $u$-plane determine the convergence radii for the Fefferman-Graham expansion. However there is a richer physical significance to the complex $u$-plane as we shall now describe.

To see this, consider real Lorentzian submanifolds of \eqref{metric} for $u \in \mathbb{C}$ and $x^\mu \in \mathbb{R}$.\footnote{We restrict our analysis to the mostly-plus signature convention, $(-,+,+,\ldots, +)$. This means that signature changes in the complex $u$-plane are not considered in our analysis, though they do occur in many of the examples we consider.}\footnote{For a related recent discussion of real submanifolds of complex metrics see \cite{Witten:2021nzp} in the context of gravitational path integrals, building on \cite{Louko:1995jw, Kontsevich:2021dmb} (see also \cite{Feldbrugge:2017kzv,DiTucci:2020weq,Lehners:2021mah, Visser:2021ucg,Jonas:2022njf}).} To determine where such submanifolds appear we take a piecewise path $\Gamma$ through the complex $u$-plane parameterised by $\lambda$ in polar coordinates, $u(\lambda) = r(\lambda) e^{i \theta(\lambda)}$ with $\lambda \in \mathbb{R}$, $r \in [0,\infty)$ and $\theta \in (-\pi,\pi]$. Then on $\Gamma$ we have the following spacetime metric,
\be
G = \left(\frac{\dot{r}^2}{r^2} + 2 i \frac{\dot{r}}{r}\dot{\theta} - \dot{\theta}^2\right) d\lambda^2+ \gamma_{\mu\nu}(u(\lambda),x)dx^\mu dx^\nu. \label{Gammametric}
\ee
Hence for $G$ to be real, it is necessary to set $2 i \dot{r}\dot{\theta}/r = 0$, which can be done in two ways, $\dot{\theta}=0$ or $\dot{r} = 0$. We conclude that real Lorentzian submanifolds of \eqref{metric} are constructed from piecewise paths $\Gamma$, where the pieces necessarily consist of:
\begin{itemize}
    \item \textbf{Rays}: $\Gamma$ for which $\dot{\theta} = 0$. The radial coordinate in the complex $u$-plane, $r$, is then a spacelike coordinate. Then for $G$ to be real Lorentzian, $\gamma_{\mu\nu}$ must also be real Lorentzian and hence $r$ labels a timelike foliation of spacetime.
    \item \textbf{Arcs} of circles centred on the origin: $\Gamma$ for which $\dot{r} = 0$. The angle in the complex $u$-plane, $\theta$, is then a timelike coordinate. Then for $G$ to be real Lorentzian, $\gamma_{\mu\nu}$ must be real Euclidean. Hence $\theta$ labels a spacelike foliation of spacetime.
\end{itemize}
From now on we use the terminology `ray' and `arc' as shorthand expressions for these submanifolds of the $u$-plane, with these specific properties, instead of in the general sense of the terms.
Note that rays correspond to spacelike geodesics at fixed $x^\mu$ (solving the geodesic equation $\dot{U}^2 = U^2$), while arcs correspond to timelike geodesics at fixed $x^\mu$ (solving the geodesic equation $\dot{U}^2 = - U^2$). 
Furthermore we can conclude that,
\begin{itemize}
    \item \textbf{intersections of arcs and rays}: correspond to bifurcation points. For example, the event horizon of an eternal black hole. 
\end{itemize}
Hence a map of the \emph{singularities}, \emph{rays} and \emph{arcs} in the complex $u$-plane paints a picture of the causal structure of the spacetime. In addition there can be further disconnected components of the boundary metric. This can then be mapped to a corresponding Penrose diagram.

\begin{figure}[h!]
\begin{center}
    \includegraphics[width=\textwidth]{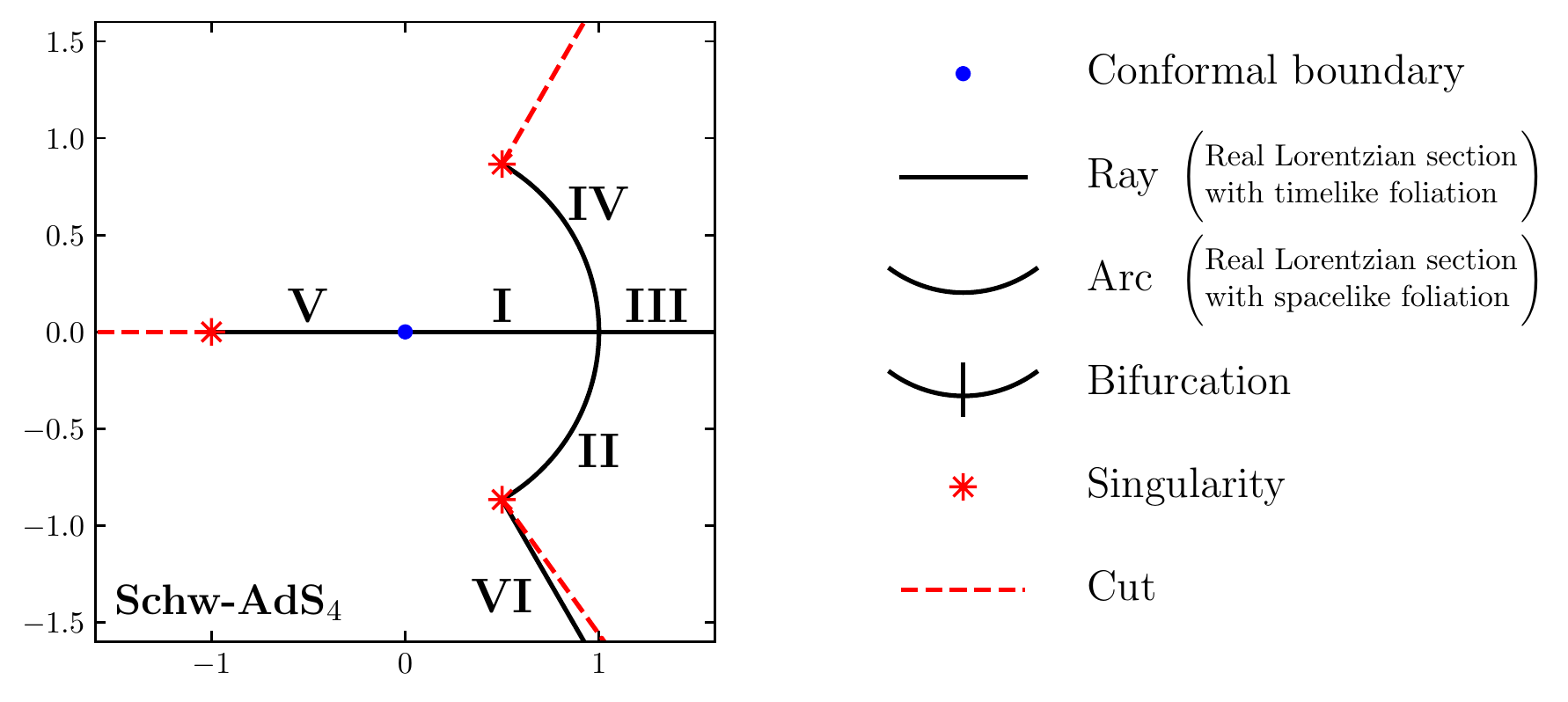}
    \caption{Planar Schwarzschild-AdS$_4$ at $\mu=2$ illustrating features that appear in the complex plane of the Fefferman-Graham coordinate, $u\in \mathbb{C}$. Here singularities refer to complex singularities of the metric components in Fefferman-Graham gauge. In this specific case these singularities are branch points which coincide with gauge-invariant curvature singularities of the maximally extended spacetime.}
    \label{fig:legend}
\end{center}
\end{figure}
\begin{figure}[h!]
\begin{center}
    \includegraphics[width=\textwidth]{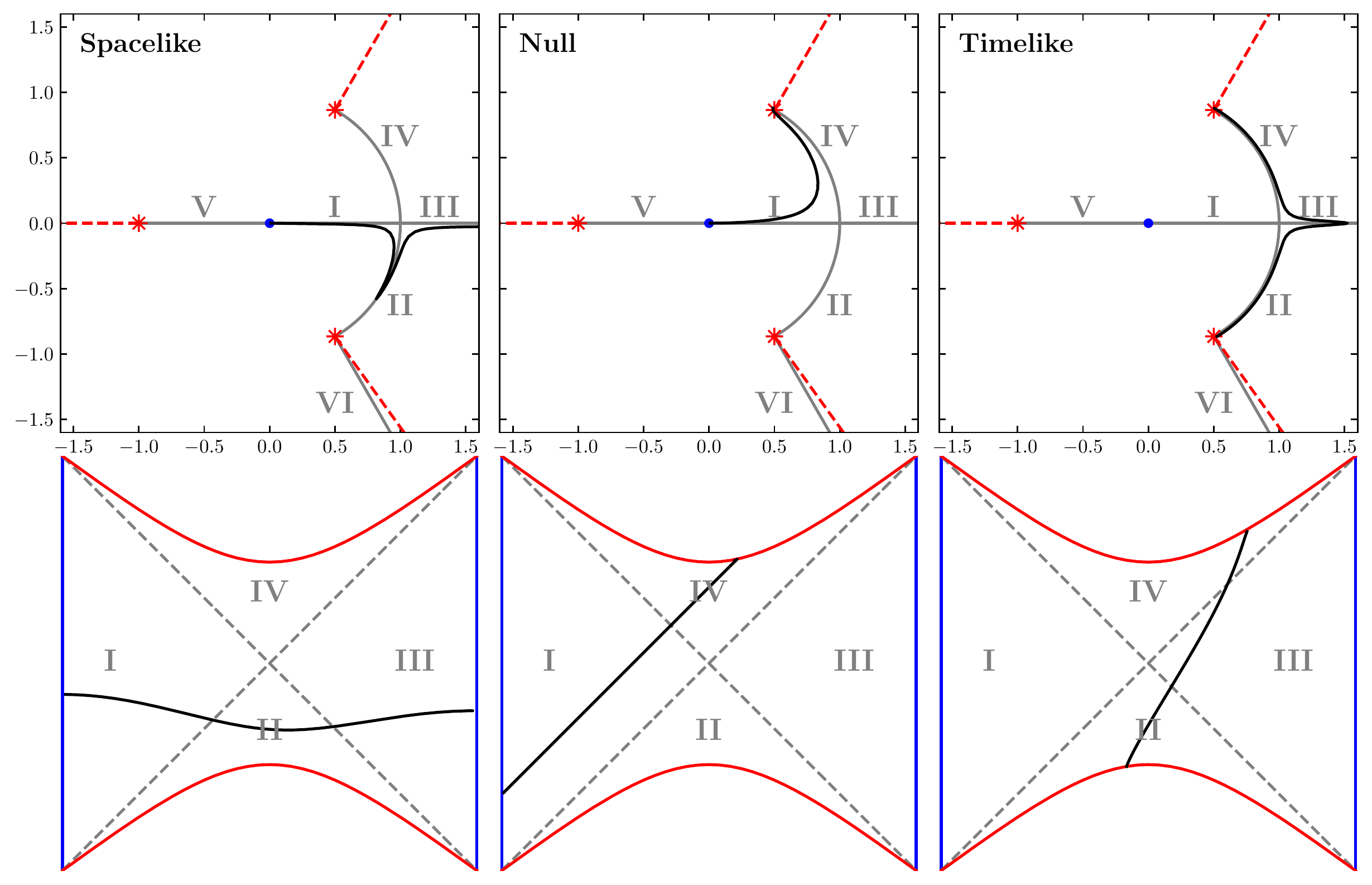}
    \caption{Selected geodesics in the complex Fefferman-Graham plane for the Schwarzschild-AdS$_4$ black brane at $\mu=2$ (black lines). \textbf{Top row:} The complex $u$-plane. For visual clarity we have added a small deviation $i\epsilon$ to the geodesic initial condition, rendering the geodesic complex. This is to be understood as visual guide to taking the limit $\epsilon \to 0$ where the geodesic becomes a piecewise path in the $u$-plane, following the rays and arcs. The spacelike geodesic reaches $u\to +\infty$ on the real $u$ line where it terminates at another portion of the conformal boundary. \textbf{Bottom row:} The corresponding $\epsilon \to 0$ geodesics plotted on the Penrose diagram, with the same colour scheme for singularities and conformal boundaries.}
    \label{fig:ex:penrose}
\end{center}
\end{figure}

Let us illustrate the physical significance of the construction of rays and arcs with an example, Schwarzschild-AdS$_4$. Details of the construction of the complex $u$-plane are given later in section \ref{sec:vacuum}, but are reproduced here in figure \ref{fig:legend} for illustrative purposes. The radius of convergence is $(2/\mu)^{1/3}$, set by branch-point singularities. These are bona fide curvature singularities of the spacetime, which can be reached by traversing physical submanifolds of the complex $u$-plane; first along the ray labelled \textbf{I} (outside the horizon) until the arc is reached (the horizon bifurcation point) and then along the arc \textbf{II} or \textbf{IV} to the singularity (inside the horizon).
Thus regions \textbf{II} and \textbf{IV} lie at the radius of convergence, while region \textbf{III} is beyond it. Region \textbf{III} connects to $u=+\infty$ corresponding to the second disconnected component of the conformal boundary in this spacetime. Region \textbf{V} along the $u<0$ axis corresponds to a Schwarzschild solution of negative mass and thus a naked timelike singularity (there is no bifurcation point on the path connecting it to the boundary). Region \textbf{VI} is a negative mass Schwarzschild solution whose conformal boundary is at infinity. The situation can be illuminated further by plotting paths taken by various geodesics in the spacetime, both on the complex $u$-plane and on the corresponding Penrose diagram. This is illustrated in figure \ref{fig:ex:penrose}. 

Based on these observations one should not be surprised to encounter situations where black holes with spacelike singularities have a radius of convergence equal to the horizon radius. This is because spacelike singularities sit on arcs, and are therefore equidistant from the origin as any bifurcation surface on the same arc. We will see this structure in many examples however it is not always the case as additional singularities can be present. For cases where the black hole has a Cauchy horizon one of the timelike black hole singularities is closer to the boundary than the bifurcation surface.

\section{Black holes in vacuum}
\label{sec:vacuum}
\subsection{Schwarzschild black brane}
\label{sec:ex:sads}
\begin{figure}[h!]
\includegraphics[width=\textwidth]{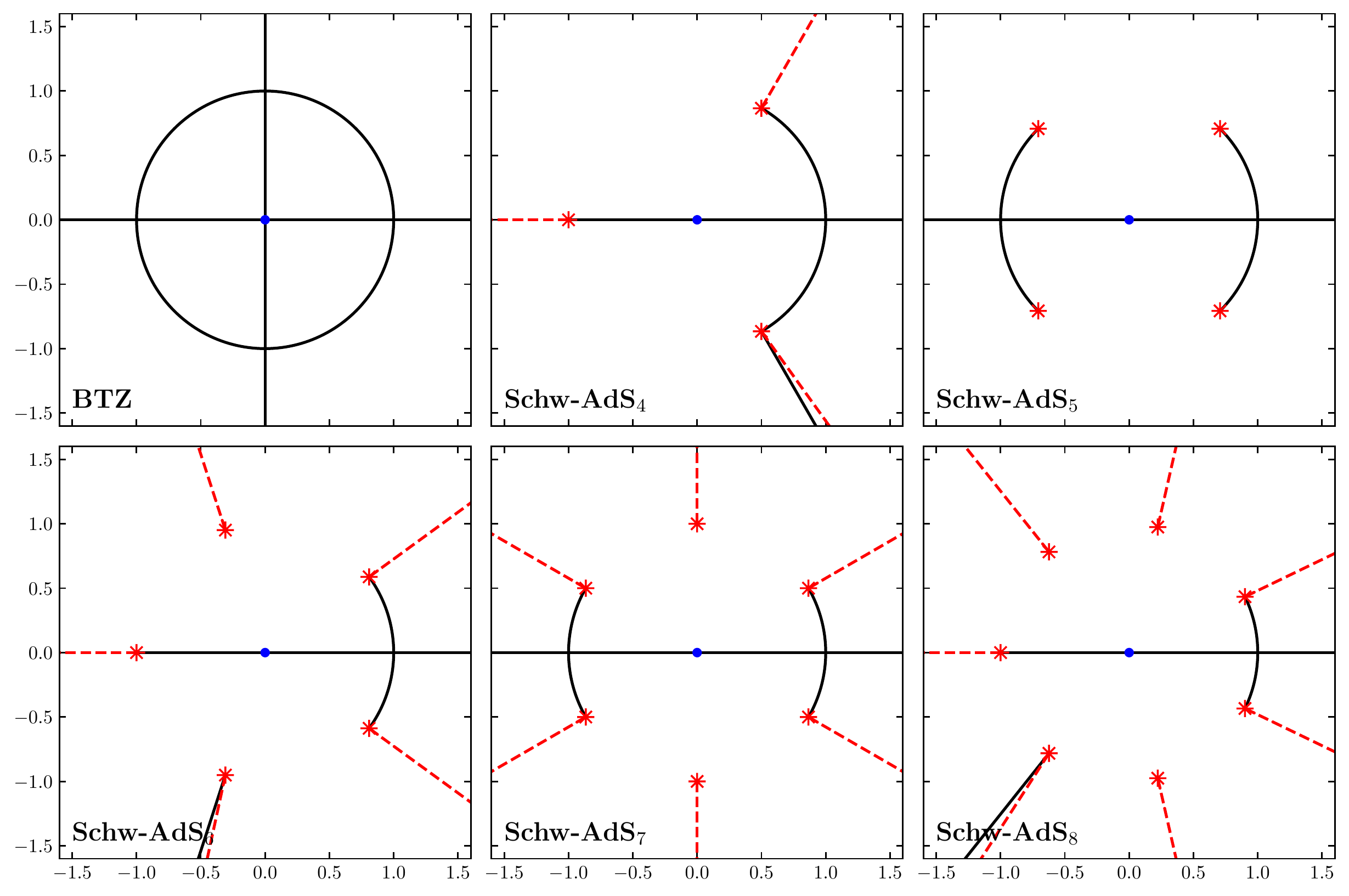}
\caption{The complex Fefferman-Graham plane for BTZ, SAdS$_4$, SAdS$_5$, SAdS$_6$, SAdS$_7$ and SAdS$_8$. The legend for these diagrams is given in figure \ref{fig:legend}.}
\label{fig:ex:psd}
\end{figure}
The Schwarzschild black brane in AdS$_{d+1}$ can be written in Schwarzschild coordinates as follows,
\be
ds^2 = \frac{1}{z^2}\left(-(1-2\mu z^d) dt^2 + \frac{dz^2}{(1-2\mu z^d)} + dx_{d-1}^2\right),
\ee
where $\mu$ is a mass parameter. The Fefferman-Graham coordinate $u$ is reached by the following coordinate transformation,
\be
z = \frac{u}{\left(1 + \frac{\mu}{2}u^d\right)^\frac{2}{d}}
\ee
bringing the metric into the form \eqref{metric}, with
\bea
\gamma &=& -f(u) dt^2 + g(u)dx_{d-1}^2,\\
f(u) &=& \frac{1}{u^2}\frac{\left(1-\frac{\mu}{2}u^d\right)^2}{\left(1+\frac{\mu}{2}u^d\right)^{\frac{2d-4}{d}}},\quad 
g(u) = \frac{1}{u^2}\left(1+\frac{\mu}{2}u^d\right)^{\frac{4}{d}}.
\eea
We have performed an exhaustive analysis of singularities, rays and arcs presented in appendix \ref{app:SAdSProofs} along with a list of results. Here we have summarised these findings for the cases $d=2,3,4,5,6,7$ in the $u$-plane plots of figure \ref{fig:ex:psd}.

\begin{figure}[h!]
\begin{center}
    \includegraphics[width=\textwidth]{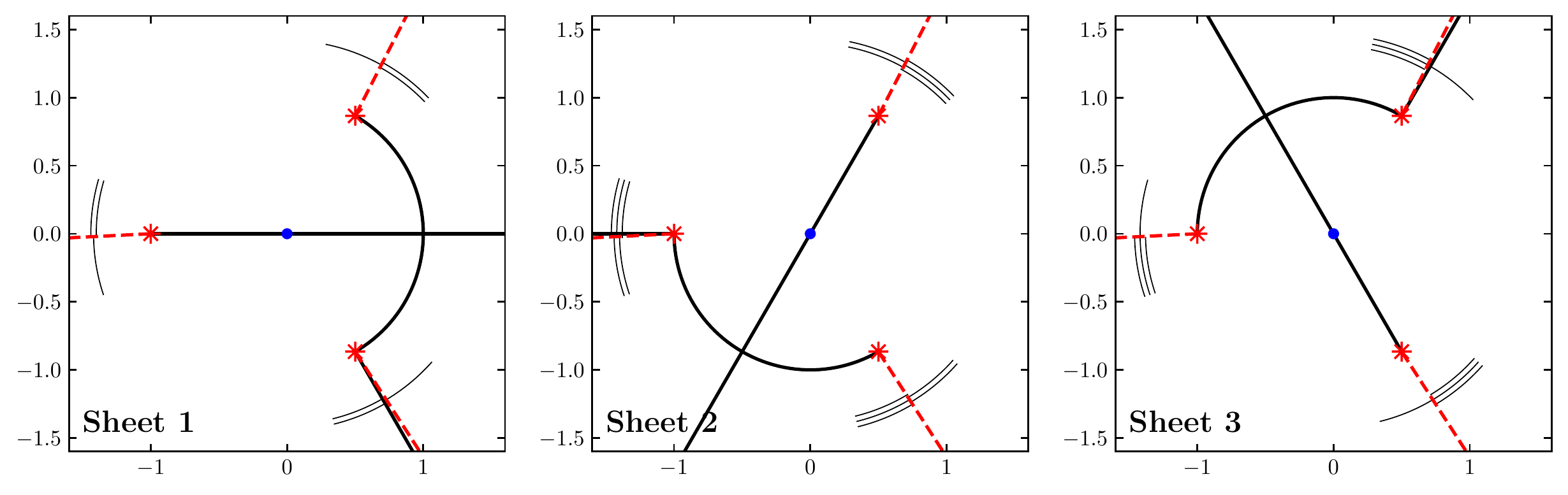}
    \caption{The three sheets of the complex Fefferman-Graham plane for the Schwarzschild-AdS$_4$ black brane. The thin black lines label the identification of sheets on each side of the indicated cuts. }
    \label{fig:ex:sads_sheets}
\end{center}
\end{figure}

For $d \geq 3$ (AdS$_4$ and above) there are singularities at $1+\frac{\mu}{2}u^d = 0$, hence $d$ singularities equally spaced around a circle of radius $(2/\mu)^\frac{1}{d}$. In $d=4$ these singularities are poles of $f$ while in all other cases they are branch points of both $f$ and $g$. These branch points mean that there are other sheets of the $u$-plane, and we have charted these in the case of $d=3$ as shown in figure \ref{fig:ex:sads_sheets}. Note however that there is no physical path which can be used to access another sheet of the $u$-plane, in contrast to other black hole spacetimes discussed later in this paper.

Along arcs, the angle in the complex plane $\theta$ is the timelike coordinate. In this case, the metric on the arcs shown is given by
\be
    ds^2 = -d\theta^2 + (2\mu)^{\frac{2}{d}}\cos^{\frac{4}{d}}\left(\frac{d}{2}\theta\right)\tan^2\left(\frac{d}{2}\theta\right)dt^2 + (2\mu)^{\frac{2}{d}}\cos^{\frac{4}{d}}\left(\frac{d}{2}\theta\right)dx_{d-1}^2.
\ee
Hence for an interior geodesic observer sitting at fixed $t$ and $x$ their proper time coordinate is $\theta$, the angle in the complex $u$-plane.
As the spacelike singularity at $\theta = \pi/d$ is approached with $\tau \to 0$ where $\theta = \pi/d-\tau$, the metric diverges in the following power-law form,
\be
    ds^2 = -d\tau^2 + \left(\frac{2}{d}\right)^\frac{4-2d}{d} (2\mu)^\frac{2}{d} \tau^{\frac{4-2d}{d}}dt^2 + \left(\frac{2}{d}\right)^\frac{4}{d} (2\mu)^\frac{2}{d} \tau^\frac{4}{d} dx_{d-1}^2 + \ldots.
\ee
In $d=3$ this is a Kasner singularity with exponents $p_1 = -1/3, p_2 = 2/3, p_3 = 2/3$ as discussed in \cite{Frenkel:2020ysx}.

For $d=2$ (BTZ) the situation is quite different. There are no metric coefficient singularities in the complex $u$-plane away from $u=0$ and the radius of convergence is infinite. This is expected since it is known that the Fefferman-Graham expansion truncates at $u^4$ when $d=2$. The $u$-plane arcs show that there are timelike geodesics permitted which orbit the origin of the $u$-plane indefinitely. Finally note that unlike $d>2$, there is a ray along the imaginary $u$-axis. This is because for $d=2$ the role of $t$ and $x$ can be exchanged, with $x$ serving as the timelike coordinate in the spacetime, enlarging the allowed set of rays. Thus four copies of the BTZ exterior spacetime meet at $u=0$, each physically accessible for a bulk observer by traversing the arcs and rays shown in figure \ref{fig:ex:psd}.

For $d$ even, there is a $u\to -u$ symmetry, that is, extending from the conformal boundary the `wrong way' along the negative real axis, one encounters the same geometry. However for $d$ odd one encounters a timelike naked singularity, as indicated by the red marker on the negative real $u$-axis. Because the singularity sits on a ray it is timelike, and because there is no bifurcation point between it and $u=0$ it is naked. This is to be expected as for odd $d$, sending $u\to -u$ corresponds to negative mass Schwarzschild, with mass $-\mu$. Perhaps less expected are the additional rays extending from a singularity to infinity. These too correspond to the Schwarzschild metric with mass $-\mu$, but whose conformal boundary is at infinity.

Finally, another discrete symmetry to consider are $u$-inversions. For all $d$, $u\to u_h^2/u$ -- where $u_h$ is the horizon radius in the $u$ coordinate -- is an isometry of the physical metric, i.e. along rays for $u>0$ and the attached arcs.  Interestingly however, for $d$ even, this isometry extends to the full complex $u$-plane, while in $d$ odd it does not. This map acts trivially in Schwarzschild coordinates, leaving $z$ invariant.

\subsection{Hawking-Hunter-Taylor with equal angular momenta}
After discussing the convergence properties of the Fefferman-Graham expansion in static black hole spacetimes, it is natural to address this question in stationary solutions of the vacuum Einstein equations with a negative cosmological constant. Rotating black holes belong to this class of spacetimes, and include the Kerr-AdS solution ($d=3$) \cite{Carter:1968ks}, its generalization to $d=4$ \cite{Hawking:1998kw} and to $d>4$ \cite{Hawking:1998kw, Gibbons:2004uw, Gibbons:2004js} (see also \cite{Emparan:2008eg} for a review).

In $d{+}1$-dimensions, a rotating black hole has $N = \lfloor\frac{d}{2}\rfloor$ independent rotation planes. For even $d$, and in the special case in which the associated $N$ angular momenta are equal, the $\mathbb R \times U(1)^N$ isometry group of the black hole gets enhanced to $\mathbb R \times U(N)$, implying that line element becomes cohomogeneity-one and depends  nontrivially on a single radial coordinate. For this reason, even $d$ equal angular momenta black holes correspond to the rotating black hole spacetimes in which the analysis of the large-order behaviour of the Fefferman-Graham expansion is the simplest from a technical standpoint. 

In this subsection, we consider the $d = 4$ case, the Hawking-Hunter-Taylor (HHT) black hole \cite{Hawking:1998kw}. The metric of the equal angular momenta HHT black hole reads 
\begin{equation}
ds^2 = - f(r)^2 dt^2 + g(r)^2 dr^2 + r^2 \hat{g}_{ab}dx^a dx^b + h(r)^2 \left(d\vartheta_3 + A_a dx^a - \Omega(r) dt \right)^2,     
\end{equation}
where 
\begin{equation}
\hat{g}_{ab}dx^a dx^b = \frac{1}{4}\left(d\vartheta_1^2 + \sin(\vartheta_1)^2 d\vartheta_2^2\right), \quad A = A_a dx^a = \frac{1}{2} \cos(\vartheta_1) d\vartheta_2,    
\end{equation}
with $\vartheta_1,\,\vartheta_2$, and $\vartheta_3$ angular coordinates on $S^3$. The functions $f$, $g$, $h$ and $\Omega$ are known in closed-form, 
\begin{subequations}\label{HHT_solution_S}
\begin{equation}
g(r) = \left(1+r^2 - \frac{2M}{r^2} + \frac{2M a^2}{r^2} + \frac{2 M a^2}{r^4} \right)^{-\frac{1}{2}}, \quad h(r) = r \left(1+ \frac{2 M a^2}{r^4}\right)^\frac{1}{2}
\end{equation}
\begin{equation}
f(r) = \frac{r}{g(r) h(r)}, \quad \Omega(r) = \frac{2 M a}{r^2 h(r)^2}. 
\end{equation}    
\end{subequations}
In equations \eqref{HHT_solution_S}, $M$ and $a$ are respectively the mass and $a$ spin parameter of the black hole. The outer (inner) horizon of the black hole is located at $r = r_+$ ($r_-$) and corresponds to the largest (smallest) positive real root of $g(r)=0$. For a given $M$, absence of a naked singularity imposes the extremality bound $a \leq a_{ext}(M)$. When the extremality bound is saturated, the inner and outer horizons become degenerate and the black hole temperature vanishes. On the other hand, for $a = 0$ the angular momenta vanish and one recovers the global Schwarzschild-AdS$_5$ black hole.\footnote{Provided the mass is positive, the complex $u$-plane representation of the Schwarzschild-AdS$_5$ black hole resembles that of the planar black hole (see figure \ref{fig:ex:psd}). The key difference is the angle of the singularities in the complex $u$-plane, and thus the proper time for a static observer to fall into that singularity. For masses $M \in (0,\infty)$ this angle takes values in the range $\theta \in (0, \pi/4)$ with $\theta = \pi/4$ as $M \to \infty$ corresponding to the planar limit. At $M=0$ the spacetime is global-AdS$_5$ and there is no singularity, but a point at finite $u>0$ where the sphere shrinks to zero, and beyond this point another copy of global-AdS$_5$ related by inversion symmetry.}

The map to Fefferman-Graham coordinates is given by the function $r(u)$ obeying the following ODE, 
\begin{equation}\label{HHT_r_eq}
r'(u) + \frac{1}{u g(r(u))} = 0.       
\end{equation}
Once that $r(u)$ is known, the metric in Fefferman-Graham coordinates is completely determined by virtue of equations \eqref{HHT_solution_S}. We are therefore interested in the large-order behaviour of the series expansion of $r(u)$ around $u = 0$,  
\begin{equation}\label{HHT_r_series}
r = r(u) = u^{-1}+\sum_{n=0}^\infty c_{2n+1} u^{2n+1},   
\end{equation}
which we determine numerically by plugging the ansatz \eqref{HHT_r_series} into equation \eqref{HHT_r_eq} and solving recursively for the series coefficients. Note that, since the functions \eqref{HHT_solution_S} are independent of the location on the $S^3$, the convergence properties of the Fefferman-Graham expansion also are, in contrast with the Kerr-AdS solution and general HHT black holes.

Our main results are two-fold: 
\begin{enumerate}
\item Irrespective of the values of $M$, $a$ and the metric component being considered, the convergence radius of the Fefferman-Graham expansion, $u_c$, is set by four equal-norm branch-point singularities in the complex $u$-plane, located at $u_s \equiv |u_s| e^{i\theta_s}$, $u_s^\star$, $-u_s$, and $-u_s^\star$,\footnote{We take $u_s$ to correspond to the singularity located in the first quadrant of the complex $u$-plane.}
\begin{equation}\label{HHT_conv_radius}
u_c = |u_s|.     
\end{equation}
\item The event horizon is located at the smallest positive real root of $r'(u) = 0$, which we denote by $u_h$. In contrast to the Schwarzschild-AdS$_5$ black brane, we now find that 
\begin{equation}
|u_s| < u_h,     
\end{equation}
implying that the Fefferman-Graham expansion stops converging before the event horizon is reached. 
\end{enumerate}

\begin{figure}[h!]
\begin{center}
\includegraphics[width=0.5\textwidth]{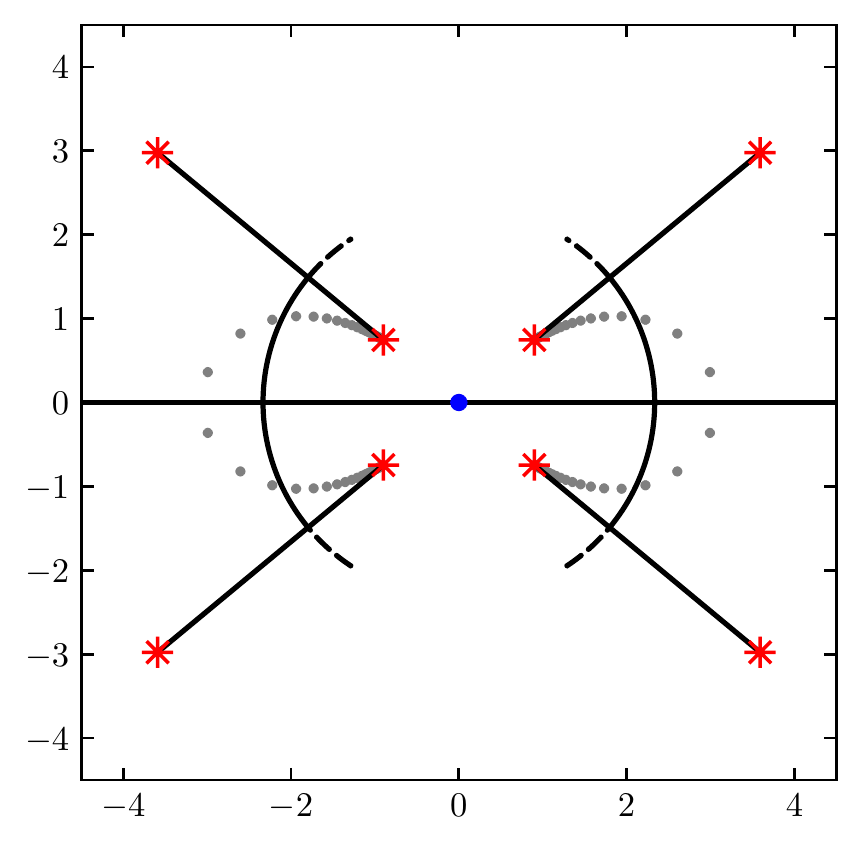}
\caption{Complex $u$-plane diagram for the equal angular momenta HHT black hole. The colour coding matches the one in figure \ref{fig:legend}. 
Grey dots correspond to the poles of a Pad\'e approximant to $r(u)$. Four lines of pole condensation are clearly visible. In the continuum limit, the poles where these lines of pole condensation start correspond to full-fledged branch-point singularities that agree with the innermost timelike singularities of the black hole. In this example, $M = 1$, $a = 0.457$, $u_h = 2.33452$, $|u_s| = 1.16828$, $|\tilde{u}_s| = 4.66495$ and $\theta_s = 0.69220$. 
} 
\label{fig:hht}
\end{center}
\end{figure}
We have cross-checked results 1. and 2. against explicit numerical solutions of equation \eqref{HHT_r_eq} as a function of complex $u$. As an additional benefit, this procedure also allows us to determine the representation of the physical spacetime in the complex $u$-plane in terms of rays, arcs and singularities. Let $u = |u| e^{i \theta} \in \mathbb C$. Our findings are summarized in the diagram shown in figure \ref{fig:hht} and discussed below. We have that:
\begin{itemize}
\item There is a ray at $\theta = 0$ extending from the left asymptotic boundary at $u = 0$ to the right asymptotic boundary at $u= \infty$. Along this ray, $r(u)$ decreases from $+\infty$ at $u=0$, hits a minimum at the outer bifurcation surface located at $u = u_h$, and then increases again to $+\infty$ at $u = +\infty$.
\item There is an arc emanating from the outer bifurcation surface at $u = u_h$, which extends from $\theta = - \theta_s$ to $\theta = +\theta_s$. Along this arc, $r(u_h e^{i\theta})$ decreases away from $\theta = 0$ until attaining a minimal value at $\theta = \pm \theta_s$. These complex-conjugated minima are new bifurcation surfaces corresponding to the inner horizons of the equal angular momenta HHT black hole. 
\item There are rays emanating from the inner horizons at $u = |u_h|e^{\pm i\theta_s}$. When moving towards smaller values of $|u|$ along these rays an endpoint is reached at $u = |u_s|e^{\pm i\theta_s}$, where $r(u)$ vanishes. Since $u_c = |u_s|$, this observation demonstrates that the branch-point singularities determining the convergence radius of the Fefferman-Graham expansion  are part of the physical spacetime and correspond to actual timelike singularities of the black hole. On the other hand, when moving towards larger values of $|u|$ along these rays, 
they do not extend indefinitely, but rather end a new branch-point singularities located at a finite value of $|u|$, $|\tilde{u}_s| = u_h^2 |u_s|^{-1}$. These new timelike singularities are at the same proper distance away from the inner bifurcation surfaces as $u_s,\,u_s^\star$, and play a role dual to them, since they determine the convergence radius of the $1/u$-expansion of $r(u)$ around the right asymptotic boundary at $u = \infty$.
\item Finally, upon increasing $|\theta|$ away from $\theta_s$, one goes through the branch cuts\footnote{For a given pair of branch-point singularities joined by a ray we always take the associated branch cut as a straight line extending from one to the other.} associated to the $u_{s}$, $u_{s}^\star$ branch points. Along the path $u = u_h e^{i\theta},\,\theta \in \mathbb R$, $r(u_h e^{i\theta})$ is a periodic function. It turns out that every maximum of this function is associated to a new outer horizon that separates a new pair of left/right asymptotic regions, while every minimum is associated to a new inner horizon that  separates a new pair of left/right timelike singularities. This is the precise way in which the maximally extended black hole spacetime is encoded in the multi-sheeted Riemann surface that emerges from our construction.  
\end{itemize}
To conclude our discussion of the equal angular momenta HHT black hole we note that, as $a \to a_{ext}(M)$ and the extremality limit is approached, $|u_s|$ tends to a finite value, while $u_h$ and $|\tilde{u}_s|$ diverge. Hence, the coordinate length of the rays emanating from the inner horizon increases without bound. 

\subsection{Kerr-AdS}

After discussing the simpler case of equal angular momenta HHT black holes, we consider the Kerr-AdS solution \cite{Carter:1968ks}. In Boyer–Lindquist coordinates, this geometry is given by  
\begin{equation}
ds^2 = - \frac{\Delta_r}{\Sigma^2} \left(dt{-} \frac{a}{\Xi}\sin(\vartheta)^2 d\phi \right)^2 + \frac{\Sigma^2}{\Delta_r} dr^2 + \frac{\Sigma^2}{\Delta_\vartheta} d\vartheta^2 + \frac{\Delta_\vartheta}{\Sigma^2}\sin(\vartheta)^2\left(a dt{-}\frac{r^2+a^2}{\Xi} d\varphi\right)^2, 
\end{equation}
where 
\begin{equation}
\Delta_r = (r^2{+}a^2)(r^2{+}1){-}2Mr,\,\,\Delta_\vartheta = 1{-}a^2 \cos(\vartheta)^2,\,\,\Sigma^2 = r^2{+}a^2 \cos(\vartheta)^2,\,\,\Xi= 1{-}a^2.    
\end{equation}
The ADM mass and angular momentum of the black hole are $M/\Xi^2$ and $J = M a/\Xi^2$.
The Kerr-AdS black hole outer (inner) horizon is located at the largest (smallest) positive real root of $\Delta_r = 0$, $r = r_+$ ($r_-$). 

The main difference between the equal angular momenta HHT black hole and the Kerr-AdS one is that, in the latter case, the map to Fefferman-Graham coordinates involves the angular coordinate $\vartheta$. One has that
\begin{equation}
r = r(u, \Theta), \quad \vartheta = \vartheta(u, \Theta),     
\end{equation}
where $r(u,\Theta)$, $\vartheta(u,\Theta)$ obey the following system of coupled PDEs, 
\begin{subequations}\label{Kerr_map}
\begin{equation}\label{Kerr_map_1}
\frac{\Sigma(r(u, \Theta), \vartheta(u, \Theta))^2}{\Delta_\vartheta(\vartheta(u,\Theta))} (\partial_u \vartheta(u, \Theta))^2  + \frac{\Sigma(r(u, \Theta), \vartheta(u, \Theta))^2}{\Delta_r(r(u,\Theta))}(\partial_u r(u,\Theta))^2 - u^{-2} = 0, 
\end{equation}
\begin{equation}\label{Kerr_map_2}
\Delta_\vartheta(\vartheta(u,\Theta)) \partial_u r(u,\Theta) \partial_\Theta r(u,\Theta) + \Delta_r(r(u,\Theta)) \partial_u \vartheta(u,\Theta) \partial_\Theta \vartheta(u,\Theta) = 0, 
\end{equation}
\end{subequations}
which correspond, respectively, to the conditions that $g_{uu} = u^{-2}$ and $g_{u\Theta} = 0$ in Fefferman-Graham coordinates. 
Equations \eqref{Kerr_map} can be solved order-by-order in a series expansion around $u = 0$, with the final result that 
\begin{subequations}
\begin{equation}\label{kerr-ads_r_exp}
r(u,\Theta) = u^{-1} + \sum_{n=1}^\infty c_n(\Theta) u^n = u^{-1} -\frac{1+a^2 \sin(\Theta)^2}{4}u + \frac{M}{3} u^2 + \ldots,  
\end{equation}
\begin{equation}
\vartheta(u,\Theta) = \Theta + \sum_{n=4}^\infty d_n(\Theta) u^n = \Theta + \frac{1}{16}a^2(-1+a^2\cos(\Theta)^2)\sin(2\Theta) +\ldots,  
\end{equation}
\end{subequations}
One finds that $d_n \propto \sin(2\Theta)$ for all $n$ and hence always vanish for $\Theta = 0, \frac{\pi}{2}$. At these particular values of the angular coordinate, $\theta(u,\Theta) = \Theta$ and $r(u,\Theta)$ decouple. We will restrict ourselves to these values, leaving a comprehensive analysis of intermediate angles for future work. 
\begin{figure}[h!]
\begin{center}
\includegraphics[width=0.5\textwidth]{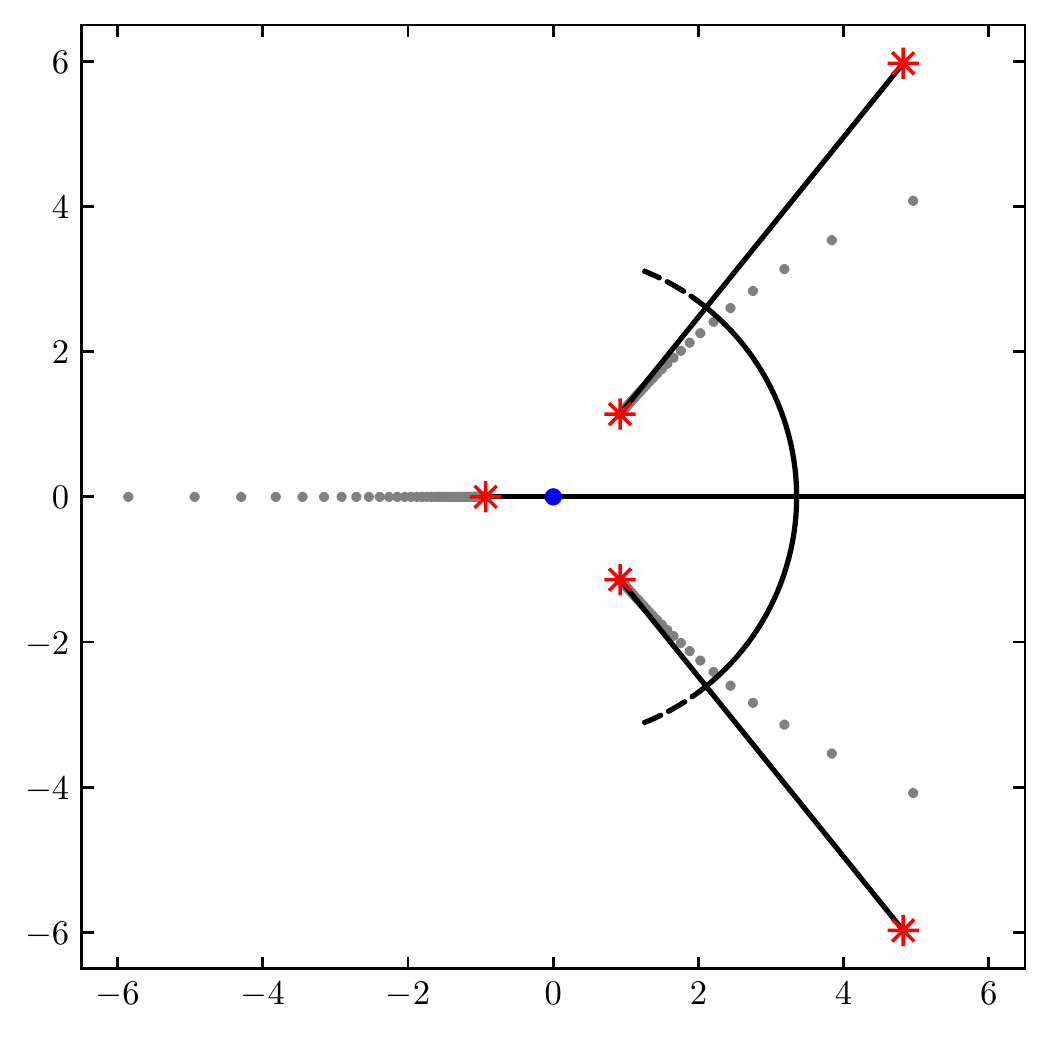}\includegraphics[width=0.338\textwidth]{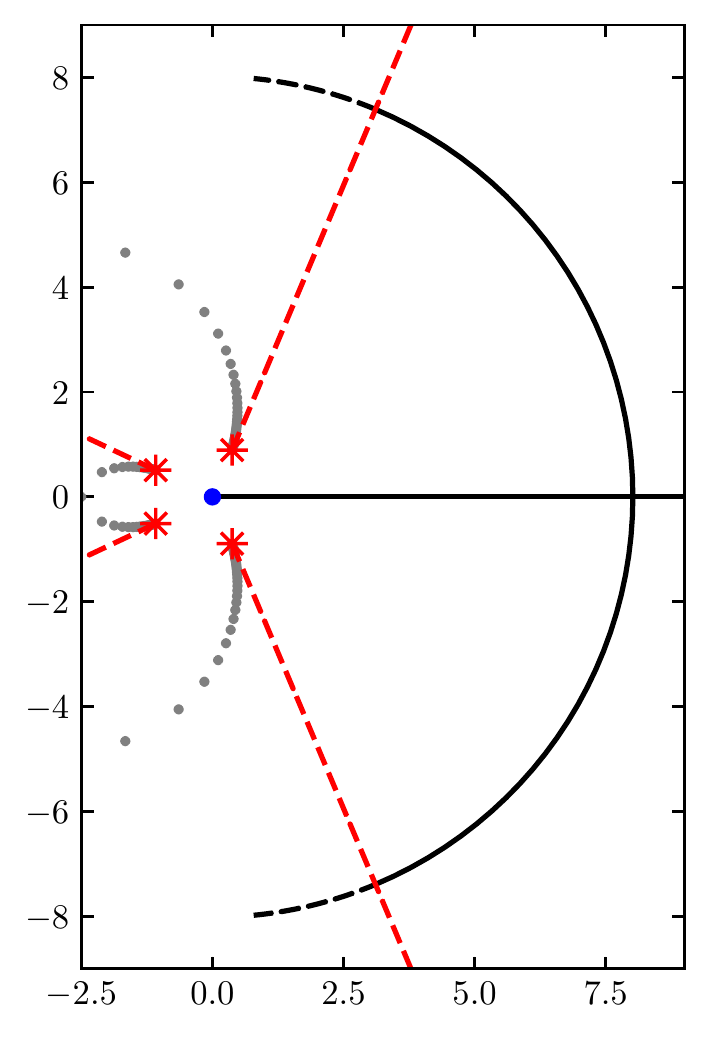}
\caption{The Kerr-AdS solution in the complex $u$-plane, for $M = 1$ and $a = 0.741$. \textbf{Left:} Equatorial plane. The correspondence between the right half-plane of the plot and the one of figure \ref{fig:hht} is manifest. Here the Padé approximant to $u~r_{eq}(u)$ shows three lines of pole condensation (grey dots) associated to three branch-point singularities, $u_{s,1}$, $u_{s,2}$ and $u_{s,2}^\star$ (red stars). The branch-point singularity with negative real part, $u_{s,1}$, corresponds to a naked singularity of the Kerr-AdS solution with reversed mass and controls the convergence radius of the Fefferman-Graham expansion. The two complex-conjugated branch-point singularities with positive real part, $u_{s,2}$ and $u_{s,2}^\star$, correspond to timelike singularities of the Kerr-AdS black hole spacetime and are accompanied by their images under inversion, $\tilde{u}_{s,2}$ and $\tilde{u}_{s,2}^\star$. Each pair of timelike singularities in the same quadrant is joined by a ray and can be associated to a branch cut running between them. For presentational clarity, the branch cuts are represented by dashed black lines that signal that the arc has entered a different sheet, rather than by dashed red lines. The parameters are $u_h = 3.3532057$, $u_{s,1}=-0.9314763$, $|u_{s,2}| = 1.464999$, $|\tilde{u}_{s,2}| = 7.675084$ and $\arg u_{s,2} = 0.8917267$. 
\textbf{Right:} Symmetry axis. The Padé approximant to $u~r_a(u)$ has four lines of pole condensation (grey dots) associated to two complex-conjugated pairs of branch-point singularities (red stars), $u_{s,1}$, $u_{s,1}^\star$ (with positive real part) and $u_{s,2}$, $u_{s,2}^\star$ (with negative real part). The former pair sets the convergence radius of the Fefferman-Graham expansion. The branch cut associated with each branch-point singularity is represented by a dashed red line. As explained in the main text, to reach the inner horizons, the arc emanating from the bifurcation surface located in the positive real axis has to go through the branch cuts associated to $u_{s,1}$, $u_{s,1}^\star$. The parameter values are $u_h = 8.0187785$, $|u_{s,1}| = 0.9686229$, $\arg u_{s,1} = 1.1725280$, $|u_{s,2}| = 1.1950925$, $\arg u_{s,2} = 2.7011671$.  
}  
\label{fig:kerr-ads}
\end{center}
\end{figure}

\subsubsection{$\Theta = \frac{\pi}{2}$}

This value of $\Theta$ corresponds to the equatorial plane of the Kerr-AdS geometry. The equation of motion for $r_{eq}(u) \equiv r(u,\frac{\pi}{2})$ simplifies to 
\begin{equation}\label{Kerr_req_eq}
r_{eq}'(u) + \frac{\left(a^2 - 2 M r_{eq}(u) + (1+a^2) r_{eq}(u)^2 + r_{eq}(u)^4 \right)^\frac{1}{2}}{u r_{eq}(u)}  = 0. \end{equation}
The induced metric on the equatorial plane is given by 
\begin{equation}
\begin{split}
&dh^2 = - h_{tt}dt^2 + \frac{du^2}{u^2} + h_{t\varphi}dtd\varphi + h_{\varphi\varphi}d\varphi^2 \\
&=-\left(r_{eq}^2+a^2+1-\frac{2M}{r_{eq}}\right)dt^2 + \frac{du^2}{u^2} + \frac{2a(-2M+a^2r_{eq}+r_{eq}^3)}{(1-a^2)r_{eq}}dtd\varphi \\
&+ \frac{2 a^2 M + (1-a^2) r_{eq}(a^2 + r_{eq}^2)}{(1-a^2)^2 r_{eq}}d\varphi^2, 
\end{split}
\end{equation}
and is completely fixed once $r_{eq}(u)$ is known. We find that the convergence radius of the small-$u$ expansion of $r_{eq}(u)$ is controlled by a branch-point singularity on the negative real axis, located at $u = u_{s,1}$, where $r_{eq}(u)$ vanishes. Analysing the lines of pole condensation of a Padé approximant to $r_{eq}(u)$ reveals the existence of two additional complex-conjugated singularities, $u_{s,2}$ and $u_{s,2}^\star$, with $|u_{s,2}| > |u_{s,1}|$ and respectively located in the first and the fourth quadrants of the complex $u$-plane. 

Solving equation \eqref{Kerr_req_eq} on the complex $u$-plane numerically corroborates the previous findings while also providing a representation of the physical spacetime in the complex $u$-plane. This representation is exactly equivalent to the equal angular momentum HHT black hole one upon replacing $u_s \to u_{s,2}$. See the left plot in figure \ref{fig:kerr-ads} for an example. The main difference between the equatorial plane of Kerr-AdS and the equal angular momenta HHT black hole appears in the left half-plane, and is due to the absence of a $u \to -u$ symmetry in the former. While in the equal angular momentum HHT solution the left half-plane is the mirror image of the right one, in the Kerr-AdS one we find that $u_{s,1}$ is connected to the origin by ray. This entails that $u_{s,1}$ corresponds to a naked singularity of the reversed mass Kerr-AdS geometry. 

\subsubsection{$\Theta = 0$}

This value of $\Theta$ corresponds to the rotation axis of the Kerr-AdS black hole. One finds that $r_a(u) \equiv r(u,0)$ is governed by 
\begin{equation}\label{Kerr_ra_eq}
r_a'(u) + \frac{(a^2 - 2 M r_a(u) + (1+a^2)r_a(u)^2 + r_a(u)^4)^\frac{1}{2}}{u (r_a(u)^2+a^2)^\frac{1}{2}}= 0,      
\end{equation}
while the induced metric is 
\begin{equation}
dh^2 = - h_{tt}dt^2 + \frac{du^2}{u^2} = - \frac{a^2 - 2 M r_a(u) + (1+a^2)r_a(u)^2 + r_a(u)^4}{r_a(u)^2+a^2}dt^2 + \frac{du^2}{u^2}. 
\end{equation}
Solving equation \eqref{Kerr_ra_eq} in a small-$u$ expansion up to a sufficiently large order and computing the corresponding Padé approximant reveals that the convergence radius of $r_a(u)$ is set by two complex-conjugated branch-point singularities, $u_{s,1}$ and $u_{s,1}^\star$, respectively located in the first and fourth quadrants of the complex $u$-plane. The Padé approximant also shows the existence of another pair of complex-conjugated branch-point singularities of greater norm, $u_{s,2}$ and $u_{s,2}^\star$, respectively located in the second and third quadrants of the complex $u$-plane.

A crucial difference with respect to the $\Theta = \frac{\pi}{2}$ case discussed previously is that these branch-point singularities are not part of the physical spacetime. This is seen by solving numerically the Einstein equations in the complex $u$-plane, from which we obtain the following representation of the physical spacetime (see the right plot in figure \ref{fig:kerr-ads}):  
\begin{itemize}
\item There is a ray extending from $u=0$ to $u=\infty$, along which $h_{tt}$ decreases from $\infty$, becomes zero at the outer horizon located at $u = u_h$, and then increases again to $\infty$. 
\item There is an arc emanating from the outer horizon. Along this arc, $r_a(u_h e^{i\theta})$ is a periodic function of $\theta$, with a maximum located at $\theta = 0$. Every new maximum (minimum) of $r_a(u_h e^{i\theta})$ corresponds to a new outer (inner) horizon. 
\item There is a ray with nonzero angle associated to every horizon that extends from $|u| = 0$ to $|u| = \infty$. Along this ray, $h_{tt}$ decreases from $\infty$, becomes zero at the horizon location, and then increases again to $\infty$. Hence, these rays are associated to new left/right asymptotic regions.  
\end{itemize}
It is important to keep in mind that, if one starts at the bifurcation surface located at $u = u_h$ and travels along the corresponding arc, reaching the inner horizons demands going through the branch cuts associated to the branch-point singularities $u_{s,1}$, $u_{s,1}^\star$. Hence, the inner horizons belong to a sheet different from the initial one. 

\section{Black holes with matter}
\label{sec:matter}
\subsection{Linear-axion black brane}
We consider the linear-axion model \cite{Andrade:2013gsa} which is an Einstein-Maxwell-scalar system admitting a closed-form black brane solution at finite charge density which explicitly breaks translation invariance. Here we consider the electrically neutral case in Schwarzschild coordinates,
\bea
ds^2 &=& \frac{-F(z)dt^2 + F(z)^{-1}dz^2 + d\vec{x}^2_{d-1}}{z^2},\\
F(z) &=& 1 - \frac{\alpha^2}{2(d-2)}z^2-m_0 z^d,\label{axionF}
\eea
together with $d-1$ scalar fields $\phi_i = \alpha x^i$.
Without loss of generality, using scaling symmetry we can set the horizon radius $z_H = 1$, then the mass parameter $m_0$ and the temperature are given by
\be
m_0 = 1 - \frac{\alpha^2}{2(d-2)}, \qquad T = \frac{1}{4\pi}\left(d - \frac{\alpha^2}{2}\right).
\ee
For future reference the Ricci scalar is
\be
R = -d(d+1) + \frac{1}{2}(d-1)z^2 \alpha^2.
\ee
$\alpha$ is the only parameter of the model. There are two interesting values, 
\be
\alpha_c = \sqrt{2(d-2)}, \qquad \alpha_\text{max} = \sqrt{2d},
\ee
for which $0 < \alpha_c < \alpha_\text{max}$. We take $\alpha$ in the range $\alpha \in \left[0, \alpha_\text{max}\right]$. The case $\alpha = \alpha_\text{max}$ is the extremal limit. At $\alpha = \alpha_c$ the temperature is positive but $m_0 = 0$. In $d=3$ at this point the holographic stress-tensor vanishes and the system has an emergent $SL(2,\mathbb{R})\times SL(2,\mathbb{R})$ duality symmetry \cite{Davison:2014lua}. From our analysis in $d=3,4$ below, we see that this point corresponds to a transition in the nature of the singularities and causal structure of the spacetime.

\subsubsection{$\alpha = \alpha_c$ for all $d>2$}
\label{sec:axion:critical}
At the critical parameter value $\alpha = \alpha_c$, where the mass parameter $m_0=0$, the map to Fefferman-Graham coordinates is given by,
\be
z = \frac{u}{1+\frac{1}{4}u^2}.
\ee
In these coordinates, the metric is
\bea
\gamma &=& -f dt^2 + g d\vec{x}^2_{d-1}\\
f &=& \frac{(4-u^2)^2}{16 u^2},\qquad g = \frac{(4+u^2)^2}{16 u^2}.
\eea
Thus we have:
\begin{itemize}
    \item \textbf{Singularities:} There are no singularities (aside from the conformal boundary).
    \item \textbf{Rays:} Rays exist for all $r>0$ excluding $r=2$, at angles $\theta =0,\pi$. 
    \item \textbf{Arcs:} Arcs exist at $r=2$ for all $\theta$ excluding the points $\theta = -\frac{\pi}{2}, 0, \frac{\pi}{2}, \pi$.
\end{itemize}
While there are no singularities of $f$ nor $g$ in the complex $u$-plane, $g$ does vanish at $u=\pm 2i$ where there is a curvature singularity of spacetime, visible as a divergence in the Ricci scalar,
\be
R = -d(d+1) + \frac{16(d-1)(d-2)u^2}{(4+u^2)^2}.
\ee
\emph{Thus in this case, the Fefferman-Graham expansion is convergent everywhere up to, including, and beyond the singularity itself.}  Indeed, in this case, the Fefferman-Graham expansion truncates at order $u^2$. This curvature singularity is accessible via an arc of the spacetime metric,
\be
ds^2 = -d\theta^2 + \sin^2\theta dt^2 + \cos^2\theta d\vec{x}_{d-1}^2,
\ee
and hence as this point is approached with $\theta = \pi/2 - \tau$ the metric has the following power-law behaviour,
\be
ds^2 = -d\tau^2 + dt^2 + \tau^2 d\vec{x}_{d-1}^2 + \ldots,
\ee
\emph{i.e.} the metric degenerates while its components are analytic there.
We note that this is not of Kasner type in $d=3$.

\subsubsection{$d=4$}
\label{sec:axion4}
\begin{figure}[h!]
\includegraphics[width=\textwidth]{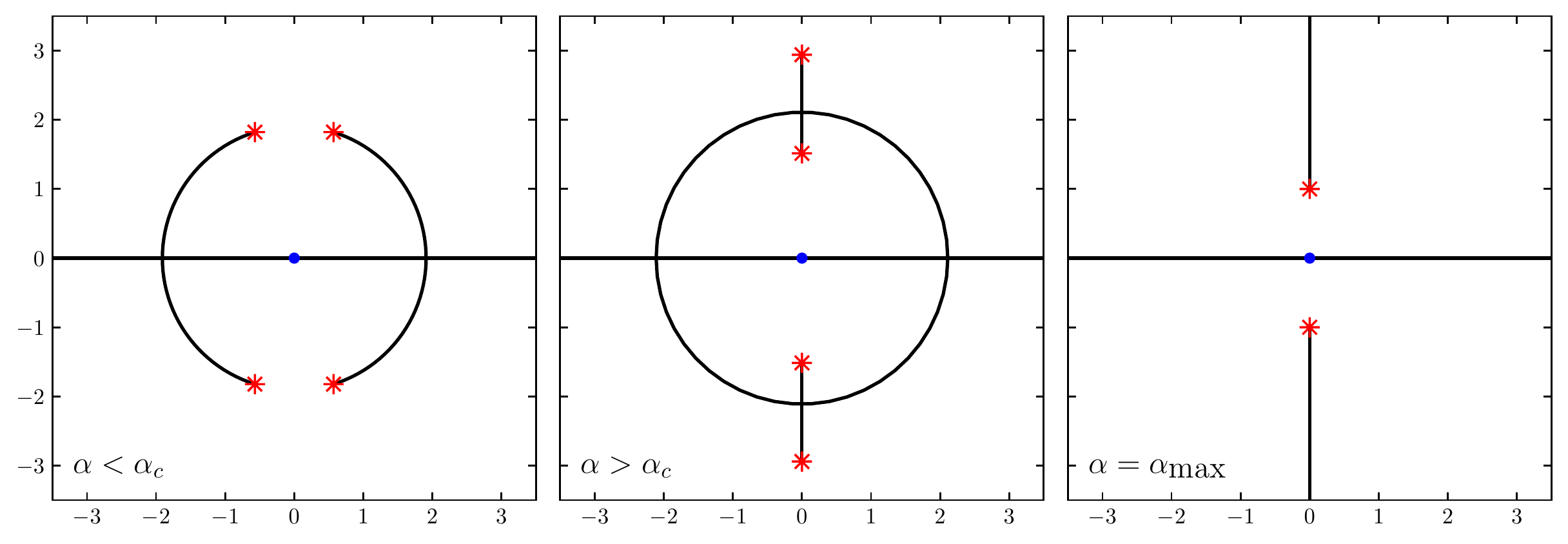}
\caption{Linear-axion black brane in $d=4$. Singularities (red), rays and arcs (black lines), conformal boundary (blue). From left to right the temperature is decreasing: $\alpha = 1.9 < \alpha_c$, $\alpha = 2.1 > \alpha_c$, $\alpha = \alpha_\text{max}$ (extremal). From panel 1 to panel 2 the singularities collide at $\alpha = \alpha_c$ and change from being spacelike to timelike. Because of this, at low enough temperatures the radius of convergence is smaller than the horizon radius.}
\label{fig:ex:axions4}
\end{figure}
\begin{figure}[h!]
\begin{center}
    \includegraphics[width=0.6\textwidth]{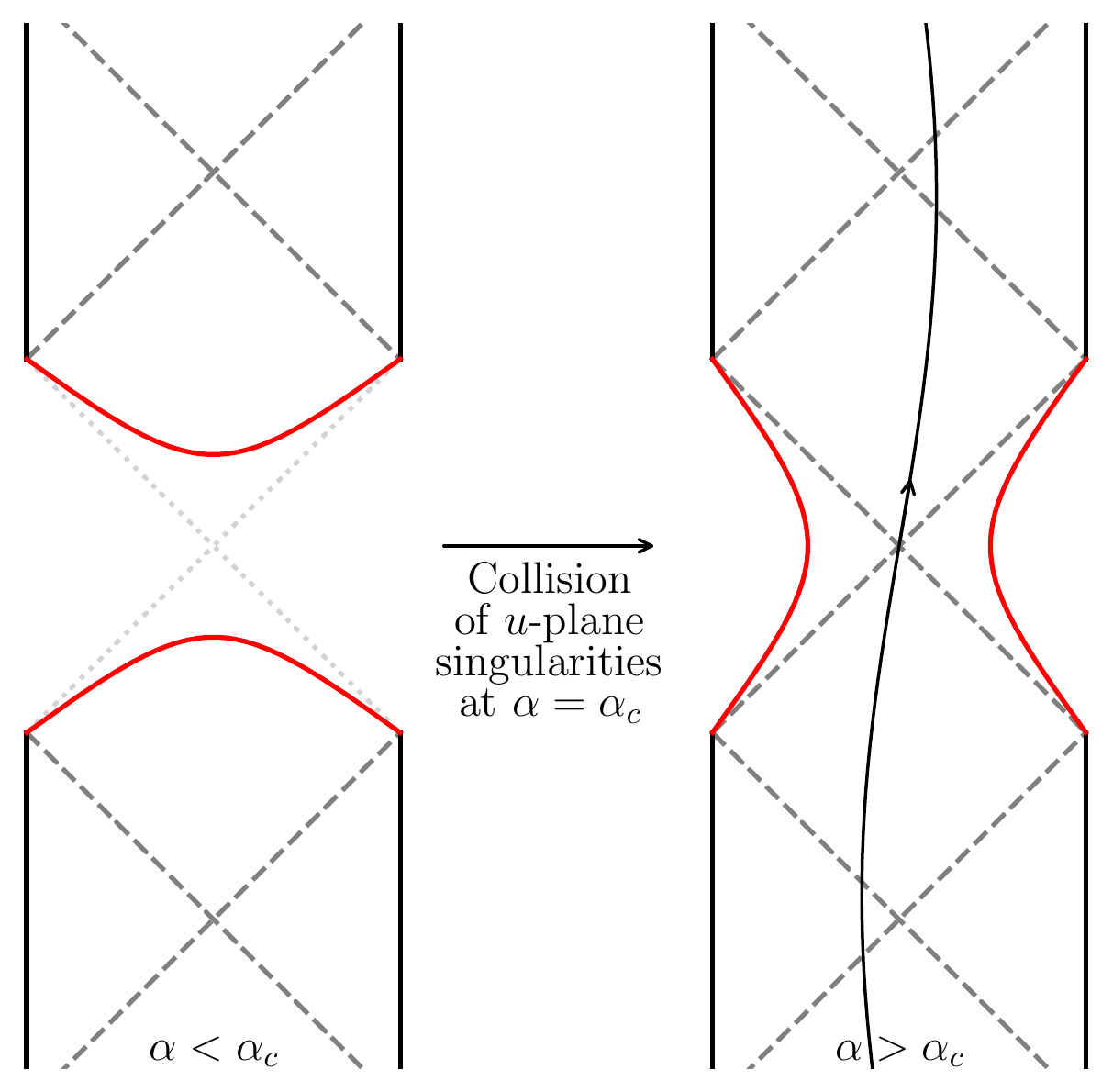}
\end{center}
\caption{Sketch of the transition in the Penrose diagram for $\alpha < \alpha_c$ to $\alpha > \alpha_c$ for the $d=4$ linear-axion model corresponding to the collision of singularities in the complex $u$-plane at $\alpha = \alpha_c$. Compare with figure \ref{fig:ex:axions4}, which shows that for $\alpha > \alpha_c$ a path opens up that allows one to orbit the origin of the $u$-plane indefinitely, passing through four bifurcation points each time, and flanked by two pairs of timelike singularities. Such a path corresponds to the existence of timelike geodesics of the sort indicated here on the right. For $\alpha < \alpha_c$ the two disconnected pieces of spacetime are simply those reached by first moving along positive $u$-axis or along the negative $u$-axis away from the conformal boundary.}
\label{fig:ex:axions4_penrose}
\end{figure}

In $d=4$ we can find the transformation to Fefferman-Graham coordinates exactly,
\be
z = \frac{16u}{\sqrt{256 + 32 \alpha^2 u^2 + (8-\alpha^2)^2 u^4}}.
\ee
In these coordinates, the metric is
\bea
\gamma &=& -f dt^2 + g d\vec{x}^2_{3},\label{axion4FG1}\\
f &=& \frac{(256-u^4(8-\alpha^2)^2)^2}{256 u^2 g_{\Sigma}},\quad g = \frac{1}{u^2} + \frac{\alpha^2}{8} + \frac{1}{256}u^2(8-\alpha^2)^2.\label{axion4FG2}
\eea
In this case the two parameter values of interest are
\be
\alpha_c = 2, \qquad \alpha_\text{max} = 2\sqrt{2}.
\ee
For this spacetime we perform an exhaustive analysis of singularities, rays and arcs, with results presented in appendix \ref{app:LA5Proofs}. The key results are summarised in figure \ref{fig:ex:axions4}. For $\alpha < \alpha_c$ there are four spacelike singularities, and for $\alpha > \alpha_c$ an interior Cauchy horizon develops and there are four timelike singularities. Then, the spacetime repeats as one orbits $u=0$ along the arcs, passing through four bifurcation points per orbit. This transition is indicated in the Penrose diagram in figure \ref{fig:ex:axions4_penrose}. In the extremal limit, the bifurcation point and outermost timelike singularities recede to infinity, but the innermost timelike singularity remains and sets the radius of convergence.

Finally, we note the behaviour of the spacelike singularities for $\alpha < \alpha_c$. Let $\theta  = \theta_\ast - \tau$, where $\theta_\ast$ is the angle of the singularity, given in appendix \ref{app:LA5Proofs}. A spacelike power-law singularity is approached as $\tau\to 0$, and governed by the following metric,
\be
ds^2 = -d\tau^2 + \frac{\sqrt{4-\alpha^2}}{5} \tau^{-1} dt^2 + \sqrt{4-\alpha^2} \tau d\vec{x}_{3}^2 + \ldots.
\ee
The case $\alpha = \alpha_c$ is discussed in the section \ref{sec:axion:critical}. When $\alpha > \alpha_c$ the proper time for a static observer to fall between horizons is always $\pi/2$.

\subsubsection{$d=3$}
\begin{figure}[h!]
\begin{center}
\includegraphics[width=\textwidth]{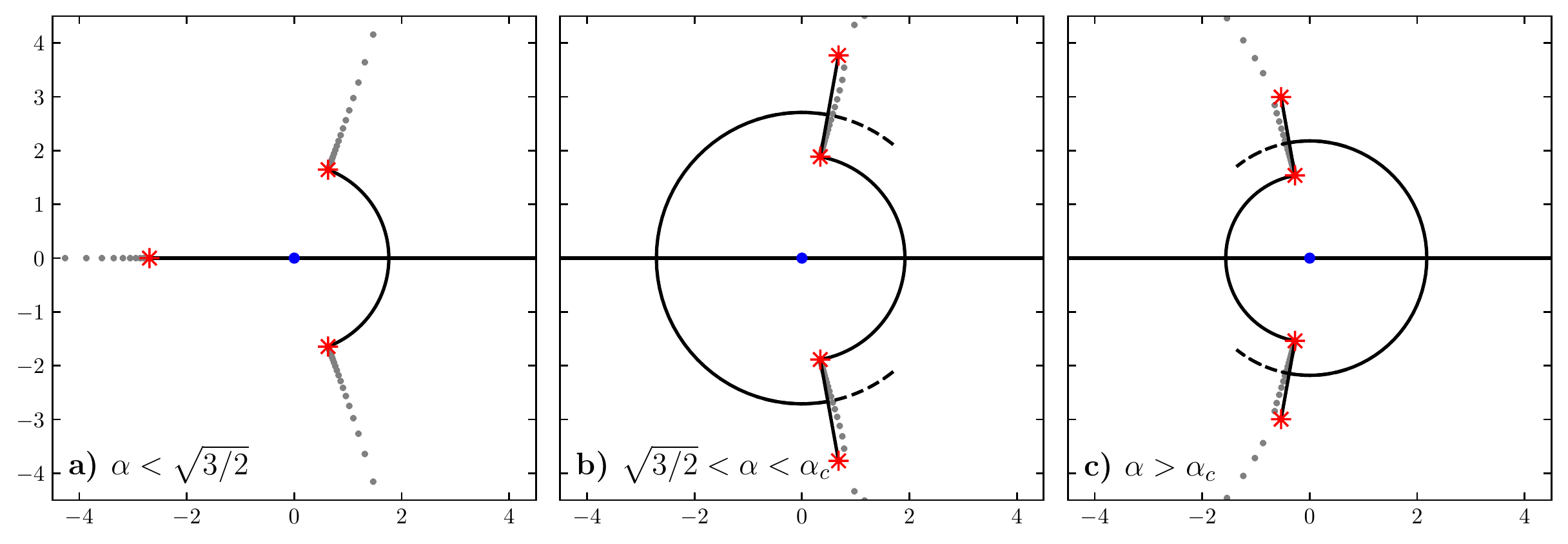}
\end{center}
\caption{The complex $u$-plane of the linear-axion black brane in $d=3$ in three qualitatively distinct regions of parameter space. The black dashed lines indicate that the arc continues past the bifurcation point onto another sheet. In a) there is a black hole with a spacelike singularity for the spacetime reached along $u>0$ and a naked singularity for $u<0$, in b) the $u<0$ spacetime develops a Cauchy horizon with arcs leading to additional sheets of the $u$-plane, while for c) the two sides exchange their causal structures. The transition from b) to c) at the critical value of $\alpha_c$ is described in section \ref{sec:axion:critical}, beyond which the radius of convergence becomes less than the horizon radius for the $u>0$ black hole.  The plot is constructed numerically and the search is not exhaustive, hence some features may be missing, e.g. no cuts are explicitly indicated. The grey dots show zeros of the Pad\'e approximant of $g$ for the Fefferman-Graham expansion, confirming the closest branch point singularities. The exact parameters shown are a) $\alpha = 1$,  b) $\alpha = 1.3$,  c) $\alpha = 1.6$.}
\label{fig:ex:axions3}
\end{figure}

When $d=3$ we cannot solve the map to Fefferman-Graham coordinates exactly. Instead we numerically construct rays, arcs and singularities. This procedure is not exhaustive. The results are shown in figure \ref{fig:ex:axions3}. 
The three distinct complex plane structures observed in figure \ref{fig:ex:axions3} can be understood as follows. Observe the similarity of the complex plane for $\sqrt{3/2} <\alpha < \alpha_c$ and $\alpha > \alpha_c$. This is due to the existence of a map relating distinct axion black holes. If $d$ is even or $\alpha = \alpha_c$ then in the original Schwarzschild coordinates, $z\to -z$ is an isometry. However consider $d$ odd and $\alpha \neq \alpha_c$, then sending $z\to -z$ gives a spacetime with the opposite mass and a different temperature.
One can show that it is a valid axion black brane with $T\geq 0$ only for certain ranges of $\alpha$. For example, when $d=3$ one can show this range is
\be
\sqrt{3/2} \leq \alpha \leq \alpha_\text{max}\qquad (d=3).
\ee
Thus a change in the complex plane structure is expected at $\alpha = \sqrt{3/2}$, which is indeed seen between panels the panels of figure \ref{fig:ex:axions3}. Another way to understand this transition is that for $\alpha < \sqrt{3/2}$ there is no second root of $F(z)$ corresponding to an inner horizon, while for $\alpha > \sqrt{3/2}$ there is.

\subsection{Reissner-Nordstr\"om black brane}
\label{subsec:RN}
 
The $d{+}1$-dimensional Reissner-Nordstr\"om (RN) black brane is dual to a thermal equilibrium state of a CFT$_d$ at finite chemical potential $\mu$. This  chemical potential is conjugate to a conserved $U(1)$ charge density $\rho$. The conserved current associated to $\rho$ is dual to an abelian gauge field in the bulk, $A = A_\mu dx^\mu$. 

In Schwarzschild coordinates, the RN black brane geometry is given by 
\begin{subequations}\label{RN_geometry}
\begin{equation}
G = - f(r) dt^2 + f(r)^{-1} dr^2 + r^2 d\vec{x}^2, 
\end{equation}
\begin{equation}
f(r) = r^2 - \left(r_h^2 + \frac{d-2}{d-1}\mu^2\right)\left(\frac{r_h}{r}\right)^{d-2} - \frac{(d-2)\mu^2}{(d-1)} \left(\frac{r_h}{r}\right)^{2d-4}, 
\end{equation}
\begin{equation}
A = A_\mu dx^\mu = \mu \left(1 - \left(\frac{r_h}{r}\right)^{d-2} \right) dt,  
\end{equation}
\end{subequations}
where $r_h$ denotes the position of the event horizon. The RN black brane temperature is non-negative provided that the extremality bound $r_h \geq r_{h,ext}=\mu (d-2)/\sqrt{d (d-1)}$ is respected.  

The map between Schwarzschild coordinates and Fefferman-Graham ones is given by the solution of the following ODE, 
\begin{equation}
\frac{r'(u)^2}{f(r(u))} - \frac{1}{u^2} = 0.        
\end{equation}
Note that $r(u)$ fixes completely $g_{tt}$, $g_{xx}$ and $A_t$ by virtue of equations \eqref{RN_geometry}. We discuss the $d = 3,\,4$ cases below. We perform our numerical computations with $\mu = \sqrt{d(d-1)}/(d-2)$, $r_{h,ext}=1$ with no loss of generality. 

\subsubsection{$d=3$}

In this case, the convergence radius of the Fefferman-Graham expansion is determined by a branch-point singularity located at $u = u_{s,1}$ on the negative real axis. Examining the lines of pole condensation of a Padé approximant to $r(u)$ reveals the existence of two additional complex-conjugated branch-point singularities located at $u = u_{s,2},\, u_{s,2}^\star$. Irrespective of the value of $T/\mu$, we always find that $|u_{s,1}| \leq |u_{s,2}|$, with equality attained in the $T/\mu \to \infty$ limit, in which the Schwarzschild black brane results are recovered. 

\begin{figure}[h!]
\begin{center}
\includegraphics[width=\textwidth]{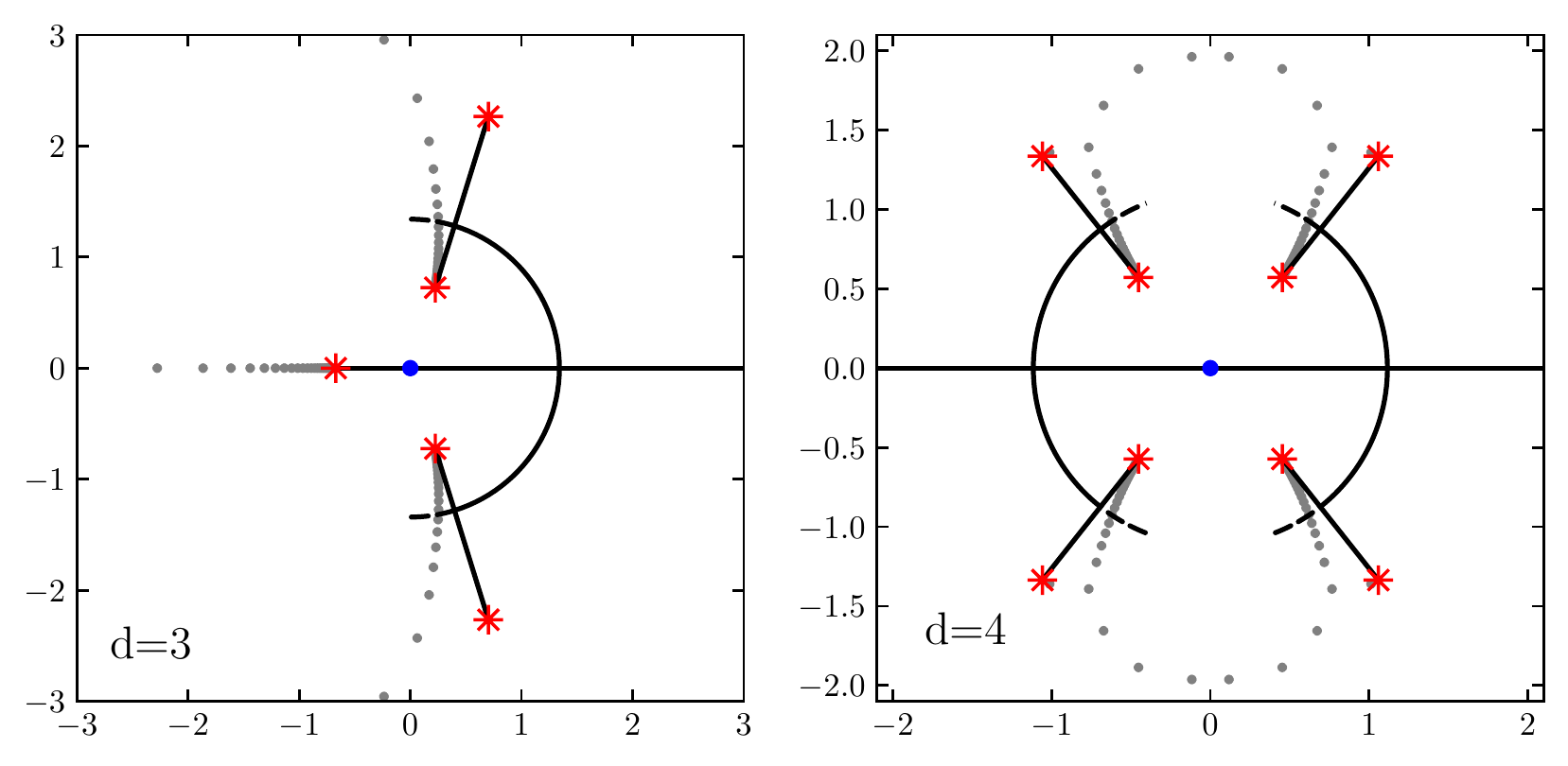}
\caption{The RN black brane in the complex $u$-plane. \textbf{Left:} $d=3$. We find five branch-point singularities (red dots), out of which three can be accessed in terms of poles of a Padé approximant (grey dots). The first branch-point singularity, $u_{s,1}$, is located on the negative real axis and despite not being part of the physical spacetime sets the convergence radius of the Fefferman-Graham expansion. The remaining four, $u_{s,2}$, $\tilde{u}_{s,2}$, $u_{s,2}^\star$ and  $\tilde{u}_{s,2}^\star$, have positive real part and correspond to actual timelike singularities of the RN black brane. They are grouped into two complex-conjugated pairs and related by inversion. We represent the branch cut extending between each pair of timelike singularities in the same quadrant by making the arc black line dashed. In this example, $\mu = \sqrt{6}$, $r_h = 1.5$, $u_h = 1.341335$, $u_{s,1} = -0.671371$, $|u_{s,2}| = 0.758480$, $|\tilde{u}_{s,2}| = 2.372083$ and $\arg u_{s,2} = 1.270263$. 
\textbf{Right:} $d=4$. The right half-plane is exactly equivalent to its $d=3$ counterpart. On the other hand, since the $d=4$ Fefferman-Graham expansion is invariant under $u \to -u$, the left half-plane is transformed into the mirror image of the right one. In this example, $\mu = \sqrt{3}$, $r_h = 1.5$, $u_h = 1.115394$,  $|u_{s,2}| = 0.730004$, $|\tilde{u}_{s,2}| = 1.704242$ and $\arg u_{s,2} = 0.901019$. 
}
\label{fig:rn}
\end{center}
\end{figure}

The physical spacetime is represented in the complex $u$-plane as follows (see the left plot in figure \ref{fig:rn}):  
\begin{itemize}
\item There is a ray at $\theta = 0$ starting at $u = 0$ and ending at $u = \infty$. $r(u)$ diverges at both ends and attains a minimum at a bifurcation surface located at $u = u_h$. This bifurcation surface corresponds to the outer horizon of the RN black brane. 
\item There is an arc emanating from $u = u_h$ and ending at $\theta = \pm \arg u_{s,2}$ on two bifurcation surfaces in which $r(u_h e^{i\theta})$ attains a minimum. These bifurcation surfaces correspond to the inner horizons to the past ($\theta = - \arg u_{s,2}$) and the future ($\theta = \arg u_{s,2}$) of the outer horizon located at $\theta = 0$. 
\item There are two rays emanating from the inner bifurcation surfaces located at $u = u_h e^{\pm i \arg u_{s,2}}$. Along each ray, $r(u)$ attains a maximum at $|u| = u_h$, and then decreases with increasing $|u-u_h|$ until it vanishes at $|u| = |u_{s,2}|$, $|u| = |\tilde{u}_{s,2}| = u_h^2/|u_{s,2}|$. These points of vanishing $r(u)$ correspond to timelike singularities of the RN black brane with distances to the origin related by inversion. As it happened in the case of the equal angular momenta HHT black hole, the member of this pair of timelike singularities located at the smallest value of $|u|$ can be accessed via Padé approximants. 
\item Upon increasing $\theta$ away from $\arg u_{s,2}$, we cross the branch cuts associated to the timelike singularities. Along this arc, $r(u_h e^{i\theta})$ is a periodic function of $\theta$. Maxima of $r(u_h e^{i\theta})$ correspond to new outer horizons separating new pairs of left/right asymptotic regions, while minima correspond to new inner horizons separating new pairs of timelike singularities.  
\end{itemize}
As a final observation, we note that $|u_{s,2}| \to u_h$ as $T/\mu \to \infty$, in such a way that the two timelike singularities associated to each inner horizon collide and become spacelike, thus recovering the Schwarzschild black brane results. In the opposite, $T/\mu \to 0$ extremal limit, both $u_h$ and the location of outermost timelike singularity of each pair diverge. 

\subsubsection{$d=4$}

The main difference between the $d = 3$ and $d=4$ RN black branes is that, in the latter case, the metric is invariant under $u \to -u$. For $\textrm{Re}~u>0$, we find the same structure as in $d=3$ while, for $\textrm{Re}~u<0$, the branch point in the negative real axis disappears and is replaced by the mirror image of the right half-plane. See the right plot in figure \ref{fig:rn}. The representation of the physical spacetime in the complex $u$-plane and the interpretation of $u_{s,2}$, $u_{s,2}^\star$, and their images under inversion as timelike singularities stays the same.

In the light of these observations, in the RNAdS$_5$ black brane the convergence radius of the Fefferman-Graham expansion is set by timelike singularities of the physical spacetime and smaller than the location of the event horizon, with equality attained only in the $T/\mu \to \infty$ limit. This puts this black hole geometry in the same universality class as the equal angular momenta HHT solution. 

\subsection{The HHH holographic superconductor}
\label{sec:HHH}
\begin{figure}[h!]
\begin{center}
\includegraphics[width=0.48\textwidth]{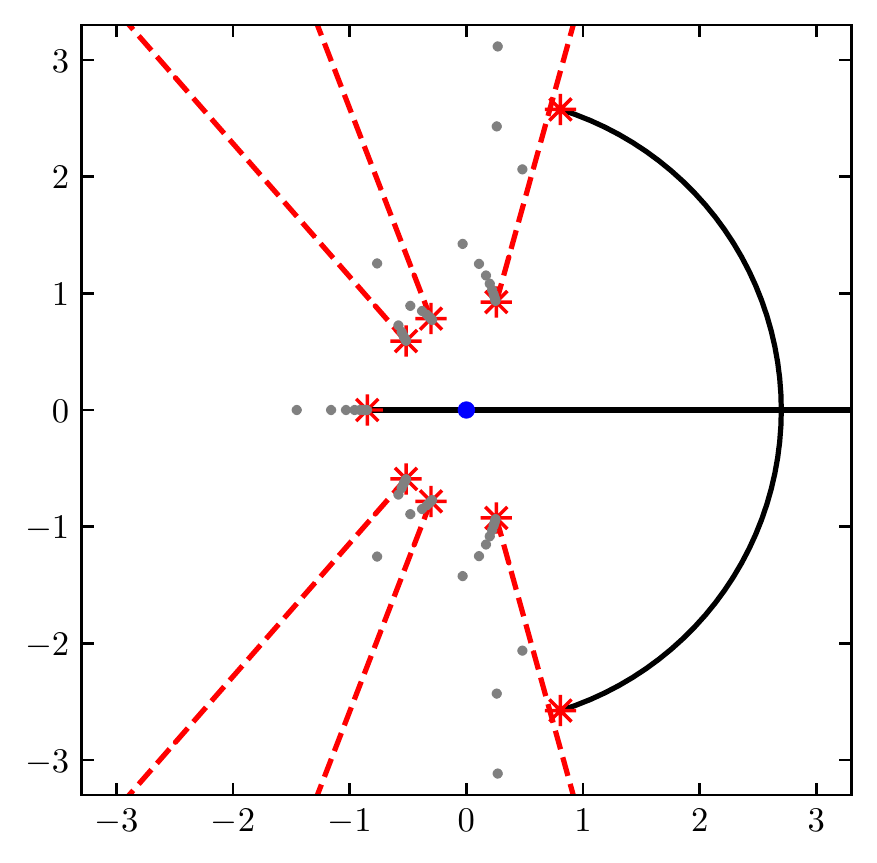}
\caption{The HHH holographic superconductor in the broken phase in the complex $u$-plane. This includes a physical extension of the spacetime into the interior and beyond the bifurcation point. Comparing to the normal phase provided by RN in figure \ref{fig:rn}, one can see that the Cauchy horizon has collapsed but a pair of former black hole singularities remain, now on a physically inaccessible portion of the $u$-plane. There are additional unphysical singularities that set the radius of convergence. The plot is constructed numerically and the search is not exhaustive, hence some features may be missing. Overlaid are zeros and poles of the diagonal Pad\'e approximant for $\psi$ (in grey).}
\label{fig:ex:HHH}
\end{center}
\end{figure}
For our final black hole example, we consider the bottom-up construction of a holographic superconductor put forward in \cite{Hartnoll:2008kx}. This is an Einstein-Maxwell-complex scalar system in $d=3$. In \cite{Hartnoll:2008kx} it is considered in Schwarzschild radial coordinate $r$, with metric functions $g, \chi$ scalar field $\psi$ and scalar gauge potential $\phi$. The bulk scalar $\psi$ is dual to a $\Delta = 2$ operator in the field theory $O_2$, while the gauge field is dual to the $U(1)$ current $J_\mu$. The system is placed at finite charge density by turning on a chemical potential $\mu$ and below a critical temperature $T_c$ the operator $O_2$ condenses with a definite phase, spontaneously breaking the $U(1)$. Without loss of generality the phase is taken to be zero so that $\psi$ is real.

The first step is to numerically construct the hairy black brane corresponding to the $U(1)$-broken phase. It is a boundary value problem governed by first order ODEs for $g,\chi$ and second order ODEs for $\psi$ and $\phi$. The boundary conditions are the conformal boundary with no sources at $r=\infty$, and a regular Killing horizon at $r=r_+$. We will not present this construction here as it is well described in detail in \cite{Hartnoll:2008kx}.  Here our only note is that we construct it with high-precision floating point numerics. Finally, converting to the Fefferman-Graham gauge near the boundary we read off the (constant) one-point functions $\left<T_{\mu\nu}\right>$,  $\left<J_\mu\right>$, and $\left<O_2\right>$. In the notation of \cite{Hartnoll:2008kx}, we take $q=1$, $\mu=3.14$, $r_+ =1$, $\Delta = 2$, $T/\mu \simeq 0.0193$.

The next step is to construct the holographic superconductor in Fefferman-Graham gauge. We take the following ansatz,
\be
\gamma = -f(u) dt^2 + g(u) (dx^2 + dy^2),\qquad \psi = \psi(u),\qquad A = \phi(u) dt.
\ee
This gives a set of ODEs, second order for $g,\psi,\phi$ and first order for $f$.
We use the CFT data from the first step, which gives us all the near-boundary data required to construct the solution in this gauge.
We construct the exact solution numerically. To do this, we integrate the bulk equations of motion along arbitrary paths in the complex $u$-plane starting at the conformal boundary at $u=0$. This allows us to determine singularities, rays and arcs in the complex $u$-plane, and hence the extension of the spacetime beyond the bifurcation point.
We also construct the Fefferman-Graham expansion to high orders, to approximately $u^{120}$. This gives us a second method of computing the radius of convergence, as well as providing additional diagnostic data using the Pad\'e approximant. The results from the two approaches are consistent with each other and are shown in figure \ref{fig:ex:HHH}. 

In the broken case there is no inner horizon, consistent with recent arguments provided in \cite{Hartnoll:2020rwq, Hartnoll:2020fhc, Cai:2020wrp}. Comparing to the unbroken phase of section \ref{subsec:RN}, the original pair of timelike black hole singularities remain in the $u$-plane, however they are now inaccessible due to the collapse of the Cauchy horizon.

\section{RG-flows: the Good, the Bad, and the Analytic}
\label{sec:rgflow}
It is possible for the Fefferman-Graham expansion to converge despite curvature singularities. A mechanism we have identified is a metric which degenerates at some radius but whose components remain analytic in Fefferman-Graham coordinates. This was seen to occur for the linear-axion model in section \ref{sec:axion:critical}. Motivated to seek further examples of this behaviour we consider the nature of singularities appearing in holographic RG flows, and their relation to the `good' singularity criterion put forward by Gubser \cite{Gubser:2000nd}. Consider the model,
\be
S = \int d^5x\sqrt{-G} \left(\frac{1}{4}R - \frac{1}{2}\left(\partial \phi\right)^2 - V(\phi)\right),
\ee
where here we restricted to a single scalar field $\phi$, together with a Poincar\'e-invariant domain wall ansatz in Fefferman-Graham coordinates,
\be
G = \frac{du^2}{u^2} + g(u)\left(-dt^2 + dx_1^2 + dx_2^2 + dx_3^2\right), \qquad \phi = \phi(u).
\ee
It is convenient to take the potential $V$ to be derived from a superpotential $W(\phi)$, so that $V = \frac{1}{8} (W')^2 - \frac{1}{3} W^2$ \cite{Skenderis:1999mm}. Then there are solutions given by the first-order equations
\be
\phi' = -\frac{1}{2u}W'(\phi),\qquad g' =\frac{2g}{3u}W(\phi).
\ee
Now consider a superpotential such that at large $\phi$ one has $W(\phi) \sim w_0 e^{\zeta \phi}$ then the infrared behaviour is governed by singularity at finite $u=u_\text{IR}$, where
\be
g  = c_2 \left(1-\frac{u}{u_\text{IR}}\right)^\frac{4}{3\zeta^2} + \ldots,\qquad \phi = -\frac{1}{\zeta}\log\left(-\frac{w_0\zeta^2}{2}\left(1-\frac{u}{u_\text{IR}}\right)\right) + \ldots
\ee
where the ellipses denote corrections in $u_\text{IR}-u$. Hence in these examples the scalar diverges logarithmically at $u_\text{IR}$, where there is a curvature singularity, $R(G)\sim (u-u_\text{IR})^{-2}$.
However, if 
\be
\zeta^2 = \frac{4}{3n}, \qquad n \in \mathbb{Z}_+ \label{theanalytic}
\ee
then the metric remains analytic at $u_\text{IR}$, despite the fact that it degenerates and has a curvature singularity. Consequently, in the absence of other singularities, the Fefferman-Graham expansion will converge up to and beyond this point for the metric. The Gubser criterion for a good singularity in this context corresponds to $\zeta^2 < 8/3$ \cite{Gubser:2000nd} and hence all of the analytic singularities outlined in \eqref{theanalytic} are of the good variety.

As a concrete example consider the superpotential
\be
W = -3 + \frac{2(\Delta -4)}{\zeta^2}\left(\cosh(\zeta \phi) - 1\right),
\ee
where $\Delta > 2$.
The system admits the following one parameter family of solutions,\footnote{In the special case where $\zeta^2 = \frac{4(4-\Delta)}{3}$ such solutions have also been found in \cite{Stanislav}, together with generalisations to all $d$.}
\be
g = \frac{1}{u^2}\left(1 - \left(\frac{u}{c_1}\right)^{8-2\Delta}\right)^\frac{4}{3\zeta^2},\qquad \phi = \frac{2}{\zeta}\,\text{arctanh}\left(\left(\frac{u}{c_1}\right)^{4-\Delta}\right),
\ee
where $c_1$ corresponds to a deformation parameter of the CFT.
Indeed, this has a branch point singularity in the metric at $u = u_\text{IR} = c_1$ unless the analyticity condition \eqref{theanalytic} holds, in which case the Fefferman-Graham expansion for the metric truncates at order $u^{(8-2\Delta)n-2}$. The truncation order can thus be arbitrarily large depending on the model parameter in the potential, $n$. The Ricci scalar diverges at $u = u_\text{IR}$ while $\phi$ has a logarithmic branch point regardless of $\zeta$. We note that the GPPZ flow \cite{GPPZ} with vanishing gaugino condensate, corresponds to precisely this solution with $\Delta = 3$ and $\zeta^2 = 4/3$, and hence obeys the analyticity condition \eqref{theanalytic}.\footnote{In the original notation of \cite{GPPZ} $u=e^{-y_\text{GPPZ}}$, $c_1 = e^{-C_{1,\text{GPPZ}}}$.}

\section{Discussion and Outlook}
\label{sec:discussion}
We have investigated the convergence properties of the Fefferman-Graham expansion, motivated by the question of how much of the bulk geometry can be constructed given a set of smooth Lorentzian CFT sources and one-point function data. For analytic data it is expected that the expansion converges in general, which we demonstrated with a number of well-known black hole and RG flow spacetimes. In these cases we concerned ourselves with finding out precisely what sets the radius of convergence through an exploration of singularities of metric functions appearing in the complex plane of the radial coordinate.

To understand the nature of these singularities we elucidated how the causal structure of the maximally developed spacetimes appears in the complex $u$-plane. For fixed boundary coordinates, $x^\mu \in \mathbb{R}$, real Lorentzian spacetime metrics are only allowed along piecewise paths made of radial rays and arcs of circles centered on the origin. Each ray and arc maps to a different portion of the associated Penrose diagram in a maximally extended spacetime. In this sense the complex $u$-plane is a useful tool to rapidly analyse the global causal structure of a black hole, where the `unit cell' of a repeating Penrose diagram may be spread over multiple sheets. In addition one may choose to take shortcuts to behind-the-horizon physics through the $u$-plane.

Our investigation of singularities proceeded by example on a case-by-case basis, so we may ask if there are any general lessons, or at least lessons common to all examples that we have considered. The first thing to note is that we have not encountered any coordinate singularities; all obstructions to convergence of the small-$u$ expansion have all been singularities in curvature invariants, or in scalar fields. For black holes with spacelike singularities in even $d$ the radius of convergence is set by the singularity of the black hole itself, and this coincides with the horizon radius of the black hole. This apparent coincidence is a natural consequence of the way that the causal structure of the spacetime is given by rays and arcs; the spacelike singularity sits on an arc, and the arc intersects a ray at the bifurcation point. Thus both are equidistant from $u=0$. For timelike singularities in even $d$, one of the pair of timelike singularities in a given region of the Penrose diagram sets the radius of convergence, and this radius is always smaller than the horizon radius.
For odd $d$ the $u$-plane is not symmetric under $u\to -u$ which creates further possibilities. For example we typically find additional singularities for $u < 0$ associated to a naked singularity of the mass-reversed spacetime. Such cases should be assessed individually.\footnote{What we have observed however is that for planar black holes the $u<0$ singularity is the same distance from $u=0$ as the black hole singularities, while for spherical black holes it is closer and for hyperbolic black holes (c.f. the axion model spacetime) it is further away.}

As we emphasised throughout, there is a crucial difference between singularities of metric components and spacetime singularities. One difference would be a coordinate singularity. Another difference is spacetime singularities with no corresponding singularity in the metric functions and thus an everywhere convergent Fefferman-Graham expansion. We have seen that this happens for a metric which degenerates in an analytic way, and provided explicit examples for both black hole spacetimes and RG flow spacetimes. In the RG flow case we were able to design a potential to have the Fefferman-Graham expansion truncate at an arbitrary order. It would be interesting to see if this can be used as a practical solution generating technique: for a given bulk system, pose an ansatz for a bulk singularity in Fefferman-Graham gauge, and tune the model parameters such that this singularity is analytic. Then in the absence of other singularities, the Fefferman-Graham expansion will truncate and thus the solution is easily constructed.  

We note that caustics do not imply a finite radius of convergence. This is easily proven with the following example. Consider global AdS$_5$ in Fefferman-Graham coordinates, $ds^2 = u^{-2}\left(du^2 - \left(1+u^2/4\right)^2 dt^2 + \left(1-u^2/4\right)^2d\Omega_3^2\right)$. The geodesic congruence generated by $u\partial_u$ form a caustic at the origin, $u=2$, with a divergent congruence expansion there, $\theta = 6(u-2)^{-1}+\ldots$. However the Fefferman-Graham expansion truncates so the radius of convergence is infinite. We also note that event horizons do not imply a finite radius of convergence, as BTZ or the critical axion model demonstrate. 

Understanding the complex $u$-plane structure for black holes allows one to decode the data appearing in the Fefferman-Graham expansion to extract information about the deep interior of the spacetime. We have illustrated this with two examples in appendices where further details can be found. The first, in appendix \ref{sec:infall_time}, discusses in detail how the large-order behaviour of the Fefferman-Graham expansion is determined by the black hole interior in the case of the RNAdS$_5$ black brane. The second, in appendix \ref{sec:conformal_map}, gives an explicit conformal map that -- for a generic class of black hole spacetimes -- maps the exterior, a portion of the interior, and the second asymptotic region, to a unit-disk in a new radial coordinate. This new radial coordinate can therefore be used to explicitly solve the Einstein equations by using a near-boundary expansion that converges all the way up to the event horizon as well as inside the black hole.

We have left many open directions which we hope to return to in future work. Of principal interest is the question of smooth but non-analytic data. There are several existence results given a smooth boundary metric \cite{GrahamLee, GURSKY2020106912, Gang}. An interesting example to consider would be a Vaidya solution where the quench is a bump function.
In such examples for locations on the boundary where the data are real analytic we expect to obtain a convergent Fefferman-Graham expansion in the sense explored in this paper, however the spacetime that it converges to would not be the global solution. For example in the Vaidya case we expect that it fails to reproduce the metric beyond the characteristic surface emanating from the non-analytic moment in time. In addition, there are CFT data for which the Fefferman-Graham expansion contains $\log u$ and higher powers thereof, and these could be handled by developing power series in both $u$ and $\log u$ as discussed in \cite{KICHENASSAMY2004268}. Finally we wish to consider time-dependent or higher cohomogeneity stationary spacetimes.\footnote{An appealing class of examples are Janus solutions. An analysis of the range of validity of the Fefferman-Graham coordinate in such a case has been made in \cite{Papadimitriou:2004rz}. It would be interesting to check if this range of validity is related to the radius of convergence.} Such series resemble a gradient expansion, and recent technical progress in analysing the large-order behaviour of both linear and nonlinear gradient expansions was made in \cite{Heller:2020uuy, Heller:2021oxl} in the context of hydrodynamic theories, which could be imported here to analyse such cases.

One may also wish to make contact with other approaches to reconstructing the bulk in AdS/CFT, especially behind the horizon. One example is the analytic continuation of CFT correlation functions \cite{Kraus:2002iv, Fidkowski:2003nf, Festuccia:2005pi, Grinberg:2020fdj}. Another is the extrapolate dictionary in which one makes operator-valued identifications between bulk and boundary \cite{Kabat:2011rz, Heemskerk:2012mn} through some kernel $K$, so that $\phi(u,x) \leftrightarrow \int K(x'|u,x)O(x') dx'$. In this vein one could explore the construction of bulk operators behind the horizon through the analytic continuation to $u \in \mathbb{C}$ that we have considered here.

Spacetimes with signatures other than the Lorentzian would also be of interest. Real submanifolds of the complex $u$-plane with different signatures arise in many of the examples we have considered in this paper. For clarity of presentation in the present Lorentzian context they have not been indicated. Some of these have multiple time directions, whilst others have mostly negative signature. Another important case is that of Euclidean AdS. For path-integral-prepared states, it is known that it is not possible to arrange an arbitrary choice of Cauchy data on a Z$_2$-symmetric slice by a suitable choice of CFT sources and vevs, as shown for probe fields in \cite{Marolf:2017kvq, Belin:2020zjb}. It would be interesting to use the techniques we have outlined here to construct a convergent nonlinear map between sources, vevs and Cauchy data in Euclidean AdS to explore this and other aspects further. 

\section*{Acknowledgements}
It is a pleasure to thank Roberto Emparan, Michal P. Heller, Christiana Pantelidou, Toby Wiseman and especially Kostas Skenderis for discussions and comments on the manuscript.
AS is supported by grant CEX2019-000918-M funded by MCIN/AEI/10.13039/501100011033. BW is supported by a Royal Society University Research Fellowship and in part by the Science and Technology Facilities Council (Consolidated Grant ST/T000775/1).

\appendix
%from here on, exclude subsections and deeper from TOC.
\addtocontents{toc}{\protect\setcounter{tocdepth}{1}}

\section{Schwarzschild black brane analysis}
\label{app:SAdSProofs}
In this section we provide an exhaustive analysis of singularities rays and arcs for the Schwarzschild black brane of section \ref{sec:ex:sads}. We also provide a summary list of results in each subsection.

As a reminder, $r \in (0,\infty)$, $\theta \in (-\pi,\pi]$ with $\mu >0$. Without loss of generality we restrict our derivation to $\mu = 2$ by scaling symmetry. A useful result in what follows is,
\be
\text{Im}(g^d) = 2(r^{2d}-1)\left(2 r^d + (1+r^{2d})\cos(d\theta)\right)\sin(d\theta). \label{img_common}
\ee

\subsection{Singularities}
If $d=2$ the metric functions $f,g$ are regular.
If $d=3$ or $d>4$ the metric functions $f,g$ have branch point singularities at $1+u^d = 0$.  If $d=4$ there are no branch points, but the denominator of $f$ has a zero at $1+ u^4 = 0$ -- the numerator is $1- u^4$ hence these correspond to poles of $f$.

To summarise, the complete list of singularities are:
\be
u_n = \left(\frac{2}{\mu}\right)^\frac{1}{d} e^{i \pi \frac{1+2n}{d}}, \quad n=0,\ldots,d-1\qquad (d\geq 3),
\ee
which are branch points except for $d=4$.

\subsection{Rays}
The condition for a ray is $f>0$ and $g>0$. 
We note that $\text{Im}(g) = 0 \implies \text{Im}(g^d) = 0$. $\text{Im}(g^d)$ is given in \eqref{img_common}; except for isolated values of $r$, this vanishes only for $\theta = n\pi/d$ with $n \in \mathbb{Z}$, i.e. $n = -d+1, \ldots, d$. Hence on a ray we find,
\be
g = \frac{(1+(-1)^n r^d)^\frac{4}{d}}{r^2} e^{-\frac{2\pi n i}{d}}.
\ee

\noindent\underline{If $n$ is even or both $n$ is odd and $r<1$} then $1+(-1)^n r^d > 0$ and hence $\text{Im}g = 0, \text{Re}g >0$ dictates that the only allowed values of $n$ are $n = 0,d$.

\noindent\underline{If $n$ is odd and $r>1$} we note $(1+(-1)^n r^d)^\frac{4}{d} = (r^d-1)^\frac{4}{d}e^{\frac{4\pi i}{d}}$. The only $n$ that satisfies $\text{Im}g = 0, \text{Re}g >0$ is $n = 2-d$ when $d$ is odd.

Hence if $d$ is even we have: $\theta = 0,\pi$. When $d$ is odd we have: $\theta =0$, $\theta = \pi$ if $r<1$, and $\theta = -\pi +  2\pi/d$ if $r>1$. One can verify that at these angles we have $f >0$ and $g>0$ except at $r=1$.

Finally, in addition to these rays, when $d=2$ a ray also occurs if $f<0$ and $g<0$. Repeating the above analysis for these conditions one finds the allowed angles are $\theta = \pm\pi/2$ for all $r>0$ excluding $r = 1$.

The complete list of rays is summarised as follows:
\begin{alignat}{3} %3 columns, 5 alignment gizmos
    \theta &= 0,   & \quad 0 < &\; r <(2/\mu)^{1/d}, && \label{ray1}\\
    \theta &= 0,   && \;r > (2/\mu)^{1/d}, &&  \label{ray2}\\
    \theta &= \pi,  & \quad 0 < &\; r <(2/\mu)^{1/d}, && \label{ray3}\\
    \theta &= \pi,    && \;r > (2/\mu)^{1/d}, && \qquad (d\text{ even})\label{ray4}\\
    \theta &= -\pi+\frac{2\pi}{d},   && \;r > (2/\mu)^{1/d}. && \qquad(d\text{ odd}) \label{ray5}\\
    \theta &= \pm \frac{\pi}{2},   & \quad 0 < &\; r <(2/\mu)^{1/d}, && \qquad(d=2)\\
    \theta &= \pm \frac{\pi}{2},   && \;r > (2/\mu)^{1/d}. && \qquad(d=2)
\end{alignat}

\subsection{Arcs}
The condition for an arc is $f<0$ and $g>0$. We note that $\text{Im}(g) = 0 \implies \text{Im}(g^d) = 0$. $\text{Im}(g^d)$ is given in \eqref{img_common}; this only vanishes at isolated values of $\theta$ or at $r=1$. Hence all arcs have $r=1$. Hence on an arc,
\be
g = (1+e^{i d \theta})^\frac{4}{d} e^{-2 i \theta},
\ee
and we note that $\arg(1+e^{2ix}) = \arctan(\tan(x)) = x - \pi \left\lfloor\frac{x}{\pi} + \frac{1}{2}\right\rfloor$. Hence,
\be
\arg(g) = - \frac{4\pi}{d} \left\lfloor\frac{d\theta}{2\pi} + \frac{1}{2}\right\rfloor, \qquad |g| = 2^\frac{2}{d} |1+\cos(d\theta)|^\frac{2}{d}.
\ee
Nonzero $|g|$ requires $\theta \neq (1+2m)\pi/d$ with $m\in \mathbb{Z}$.
Positivity requires $\left\lfloor\frac{d\theta}{2\pi} + \frac{1}{2}\right\rfloor = \frac{n d}{2}$ for $n\in \mathbb{Z}$. 

\noindent\underline{For $d$ even} with $\theta \in (-\pi,\pi]$ we have $n=-1,0,1$ and taking into account $|g|$ this corresponds to the intervals $-\pi < \theta < -\pi + \pi/d$, $-\pi/d < \theta < \pi/d$ and $\pi - \pi/d < \theta \leq \pi$.

\noindent\underline{For $d$ odd} $n$ must be even, but considering the range of $\theta$ we can only have $n=0$, and hence $-\pi/d < \theta < \pi/d$.

Finally we must check these intervals for $f$, we have
\be
\frac{f}{g} = - \tan^2\left(\frac{d\theta}{2}\right)
\ee
and hence since $g > 0$ this ensures that $f$ is manifestly real and non-positive. All we need to do is ensure that $f\neq 0$. Hence $\theta \neq  q 2\pi/d$ with $q \in \mathbb{Z}$. This removes the points $\theta = 0,\pi$ from the four intervals above.

Thus the complete list of arcs is as follows:
\begin{alignat}{3}
r &= (2/\mu)^\frac{1}{d}, &\quad 0<&\;|\theta| < \pi/d, &&\\
r &= (2/\mu)^\frac{1}{d}, &\quad \pi-\pi/d<&\; |\theta| < \pi.&& \qquad(d\text{ even})
\end{alignat}

\section{$d=4$ Linear axion black brane analysis}
\label{app:LA5Proofs}
In this section we provide a derivation of all singularities, rays and arcs for the black branes of section \ref{sec:axion4}.
For convenience let us denote,
\be
r_H = \frac{4}{\sqrt{8- \alpha^2}},\qquad \delta \theta = \frac{1}{2}\arccos\left(\frac{\alpha^2}{8-\alpha^2}\right).
\ee
Consider the metric in Fefferman-Graham gauge, \eqref{axion4FG1} and \eqref{axion4FG2}. In polar coordinates, 
\be
g = \frac{\alpha^2}{8} + \left(\frac{(\alpha^2-8)^2}{256}r^2+\frac{1}{r^2}\right)\cos(2\theta) + i \left(\frac{(\alpha^2-8)^2}{256}r^2-\frac{1}{r^2}\right)\sin(2\theta).
\ee
As a reminder, $r \in (0,\infty)$, $\theta \in (-\pi,\pi]$ and $\alpha \in [0,2\sqrt{2}]$.

\subsection{Singularities}
The metric contains no branch points, and so the only singularities aside from $u=0$ are poles of $f$. These occur at $g = 0$ provided the numerator $256-u^4(8-\alpha^2)^2 \neq 0$ there. 

\noindent\underline{For $\theta \neq n \pi/2$ where $n\in \mathbb{Z}$} then this can only occur at $r = r_H$. This leaves a real
\be
g = \frac{\alpha^2}{8} + \frac{8-\alpha^2}{8}\cos(2\theta), \label{a4proof_gsigma}
\ee
which can only vanish if $0\leq \alpha < 2$ (at $\alpha =2$ it vanishes at $\theta = n\pi/2$ and thus excluded here) and 
\be
\theta \in \left\{-\frac{\pi}{2} - \delta\theta,-\frac{\pi}{2} + \delta\theta,\frac{\pi}{2} - \delta\theta,\frac{\pi}{2} + \delta\theta\right\}. \label{a4proof_theta}
\ee
Finally one can verify that these are all poles of $f$.

\noindent\underline{For $\theta = n \pi/2$ where $n = 2m$ is even} with $m \in \mathbb{Z}$ then
\be
g = \frac{(\alpha^2-8)^2}{256}r^2+\frac{1}{r^2} + \frac{\alpha^2}{8},
\ee
each term is manifestly positive and thus $g$ cannot vanish. 

\noindent\underline{For $\theta = n \pi/2$ where $n = 2m+1$ is odd} with $m \in \mathbb{Z}$ then
\be
g = -\frac{(\alpha^2-8)^2}{256}r^2 - \frac{1}{r^2} + \frac{\alpha^2}{8}.
\ee
In this case the only points for which $g = 0$ are then
\be
r = \frac{4\left(2\pm\sqrt{\alpha^2-4}\right)}{8-\alpha^2},
\ee
with $\alpha \geq 2$, with the exception of the case $\alpha = 2\sqrt{2}$ where there is only the $(-)$ root giving $r=1$.
Finally we verify that these are poles of $f$, except when $\alpha = 2$. 
Hence a final listing of singularities is as follows:
\begin{alignat}{3}
 r &= r_H,  &\quad |\theta| &= \frac{\pi}{2} \pm \delta\theta, &\quad (0 \leq \alpha < 2)&\\
 r &= \frac{4\left(2\pm\sqrt{\alpha^2-4}\right)}{8-\alpha^2}, &\quad |\theta| &= \frac{\pi}{2}, &\quad (2 < \alpha < 2\sqrt{2})&\\
  r &= 1, &\quad |\theta| &= \frac{\pi}{2}. &\quad ( \alpha = 2\sqrt{2})&
\end{alignat}

\subsection{Rays}
Rays require $f>0$ and $g > 0$. The imaginary part of $g$ only vanishes along $\theta = n\pi/2$ with $n\in \mathbb{Z}$ or at isolated points in $r$. 

\noindent\underline{For $\theta = n\pi/2$ where $n=2m$ is even} with $m\in \mathbb{Z}$ one has
\be
f = \frac{(256-r^4(8-\alpha^2)^2)^2}{256 r^2 (256 + 32 \alpha^2 r^2 + (8-\alpha^2)^2 r^4)}, \quad g = \frac{256 + 32 \alpha^2 r^2 + (8-\alpha^2)^2 r^4}{256r^2},
\ee
and hence $f > 0$ and $g >0$ for all $r \in (0,\infty) \setminus \{r_H\}$. 

\noindent\underline{For $\theta = n\pi/2$ where $n=2m+1$ is odd} with $m\in \mathbb{Z}$ one has
\be
f = \frac{(256-r^4(8-\alpha^2)^2)^2}{256 r^2 (-256 + 32 \alpha^2 r^2 - (8-\alpha^2)^2 r^4)}, \quad g = \frac{-256 + 32 \alpha^2 r^2 - (8-\alpha^2)^2 r^4}{256r^2},
\ee
for $f > 0$ and $g >0$ we therefore require
\be
-256 + 32 \alpha^2 r^2 - (8-\alpha^2)^2 r^4 > 0, \quad \text{and} \quad r \neq r_H. \label{a4proof_raycond}
\ee
Condition \eqref{a4proof_raycond} is only satisfied if $\alpha = 2\sqrt{2}$ and $r \in (0,\infty)\setminus\{r_H\}$, or if $2 < \alpha < 2\sqrt{2}$ and
\be
r \in \left(\frac{4(2-\sqrt{\alpha^2 - 4})}{8-\alpha^2}, \frac{4(2+\sqrt{\alpha^2 - 4})}{8-\alpha^2}\right) \setminus \left\{ r_H \right\}.
\ee
A final listing of rays is as follows: 
\begin{alignat}{3}
\theta &= 0, &\quad  0 < \;& r < r_H,  &\quad (0\leq \; & \alpha \leq 2\sqrt{2})\\
\theta &= 0, &\quad  \;& r > r_H, &\quad (0\leq \; & \alpha \leq 2\sqrt{2})\\
\theta &= \pi, &\quad  0 < \;& r < r_H,  &\quad (0\leq \; & \alpha \leq 2\sqrt{2})\\
\theta &= \pi, &\quad  \;& r > r_H, &\quad (0\leq \; & \alpha \leq 2\sqrt{2})\\
|\theta| &= \frac{\pi}{2}, &\quad \frac{4(2-\sqrt{\alpha^2 - 4})}{8-\alpha^2} < \;& r < r_H, &\quad (2 < \;& \alpha < 2\sqrt{2})\\
|\theta| &= \frac{\pi}{2}, &\quad r_H < \;& r < \frac{4(2+\sqrt{\alpha^2 - 4})}{8-\alpha^2}, &\quad (2 < \;& \alpha < 2\sqrt{2})\\
|\theta| &= \frac{\pi}{2}, &\quad & r>1. &\quad (&\alpha = 2\sqrt{2})
\end{alignat}

\subsection{Arcs}
Along arcs we must have $f < 0$ and $g >0$. The reality condition for $g$ is $r=r_H$ up to isolated points in $\theta$, then on this circle,
\be
f = -\frac{(8-\alpha^2)^2 \sin^2(2\theta)}{8\left(\alpha^2 + (8-\alpha^2) \cos(2\theta)\right)},\qquad g = \frac{1}{8} \left(\alpha^2 + (8-\alpha^2) \cos(2\theta)\right).
\ee
Hence the condition for an arc is
\be
\alpha^2 + (8-\alpha^2) \cos(2\theta) > 0, \quad \text{and} \quad \theta \neq \frac{n\pi}{2}\qquad n\in \mathbb{Z}. \label{a4proof_arccond}
\ee
When $2 < \alpha \leq 2\sqrt{2}$ the first condition is always satisfied, and hence in this case we have arcs for $\theta \in (-\pi,\pi]\setminus\{-\pi/2,0,\pi/2,\pi\}$. For the remaining parameter range, $0\leq \alpha \leq 2$ we note that the intervals in $\theta$ for which \eqref{a4proof_arccond} hold are delineated by the values of $\theta$ given in \eqref{a4proof_theta}. We just need to check the signs on each side of these points to identify the admissible regions. These are,
\be
\left\{0 < \theta < \frac{\pi}{2} - \delta\theta,\;\;
\frac{\pi}{2} + \delta\theta < \theta < \pi,\;\;
-\frac{\pi}{2} + \delta\theta < \theta < 0,\;\;
-\pi < \theta < -\frac{\pi}{2} - \delta\theta\right\}.
\ee
A final listing of arcs is as follows:
\begin{alignat}{3}
    r &= r_H, &\quad 0 < \;& |\theta| < \frac{\pi}{2} - \delta\theta, &\quad (0\leq \;&\alpha \leq 2)\\
    r &= r_H, &\quad \frac{\pi}{2} + \delta\theta < \;&|\theta| < \pi, &\quad (0\leq \;&\alpha \leq 2)\\
    r &= r_H, &\quad 0 < \;&|\theta| < \frac{\pi}{2}, &\quad (2 < \;&\alpha < 2\sqrt{2})\\
    r &= r_H, &\quad \frac{\pi}{2} < \;&|\theta| < \pi. &\quad (2 < \;&\alpha < 2\sqrt{2})
\end{alignat}

\section{Behind-the-horizon physics from asymptotic behaviour}\label{sec:infall_time}

In the main body of the paper, we have studied several examples of black hole spacetimes in which the convergence radius of the Fefferman-Graham expansion was determined by the location of the black hole singularities in the complex $u$-plane. This is a particular instance of a general relationship between the asymptotic behaviour of the Fefferman-Graham expansion and the behaviour of the metric components in the vicinity of the singularities, 
\begin{equation}\label{relationship}
\textrm{Near-singularity behaviour} \longleftrightarrow \textrm{Large-order behaviour} 
\end{equation}
In this appendix, we explore the general relationship \eqref{relationship} in detail, focusing on the left-to-right arrow. The mathematical tools we will utilize are Darboux theorem and generalizations thereof \cite{szego, wilf2005generatingfunctionology}. Our analysis will demonstrate that, while the Fefferman-Graham expansion is constructed solely from boundary CFT sources and one-point functions, its large-order behaviour alone can provide a window into behind-the-horizon physics.\footnote{By this, we mean that behind-the-horizon physics can be accessed \emph{without} having to analytically continue the Fefferman-Graham expansion or finite-order truncations thereof, for instance by utilizing Padé approximants.}

Let us assume that the black hole spacetime we are interested in is endowed with $p$ equal-norm, closest-to-the-origin branch-point singularities in the complex $u$-plane located at $u = u_s^{(k)},~k\in\{1,\ldots,p\}$. Szeg\"o theorem \cite{szego, wilf2005generatingfunctionology} states that any function $f(u)$ analytic in $|u|<|u_s^{(k)}|$ with this singularity structure is such that %There is a typo in Szego and Wilf.
\begin{equation}\label{szego}
[u^n](f) \sim  \sum_{\nu = 0}^\infty\sum_{k=1}^p c_\nu^{(k)} (-1)^n {{\alpha^{(k)} + \nu \beta^{(k)}}\choose{n}}\left(u_s^{(k)}\right)^{-n},     
\end{equation}
where $c_\nu^{(k)}$, $\alpha^{(k)}$ and $\beta^{(k)}>0$ are defined in terms of the Puiseux series of $f(u)$ around $u = u_s^{(k)}$, 
\begin{equation}
f(u) = \sum_{\nu=0}^\infty c_\nu^{(k)} \left(1-\frac{u}{u_s^{(k)}}\right)^{\alpha^{(k)}+ \nu \beta^{(k)}},  \end{equation}
and $[u^n](f)$ is the $n$-th order coefficient of the series expansion of $f(u)$ around $u=0$. 

In the case that the black hole singularities determining the convergence radius of the Fefferman-Graham expansion were poles (as in the Schwarzschild-AdS$_5$ or the $d=4$ linear-axion black branes) there are analogous theorems one can resort to (see e.g. theorem 5.5 in \cite{wilf2005generatingfunctionology}).
\\\\
\noindent As an example of application of Szeg\"o theorem, we focus on the RNAdS$_5$ black brane. Furthermore, we restrict ourselves to the leading-order $\nu = 0$ contribution to \eqref{szego}, and hence drop the subscript in $c^{(k)}_0$ to avoid clutter. As we demonstrated in section \ref{subsec:RN}, the RNAdS$_5$ black brane is endowed with four branch-point singularities located at $u_s^{(1)} = u_s$, $u_s^{(2)}=-u_s^\star$, $u_s^{(3)}=-u_s$, $u_s^{(4)}=u_s^\star$ that set the convergence radius of the Fefferman-Graham expansion. These singularities are part of the physical spacetime and timelike. The function we focus on is $B = u^2 g_{xx}(u)$. 

At $u = u_s^{(k)}$, $g_{xx}$ and $B$ vanish. Solving the Einstein equations around this point, one finds that $\alpha = \frac{2}{3}$ and that $c^{(k)}$ is a root of the following cubic equation, 
\begin{equation}
(c^{(k)})^3 = 6 r_h^4 \mu^2 (u_s^{(k)})^6.     
\end{equation}
The branch $c^{(k)}$ belongs to has be chosen carefully. It is selected by the following reality conditions, 
\begin{itemize}
\item For $k = 1$, $g_{xx}(u_s(1+\epsilon e^{i 0^-}))$ is real for $\epsilon \geq 0$, implying that $c^{(1)} = 6^\frac{1}{3}e^{-\frac{2\pi i}{3}} (u_s^{(1)})^2 r_h^\frac{4}{3} \mu^\frac{2}{3}$. 
\item For $k =2$, $g_{xx}(-u_s^\star(1+\epsilon e^{i 0^+}))$ is real for $\epsilon \geq 0$, implying that $c^{(2)} = 6^\frac{1}{3}e^{\frac{2\pi i}{3}} (u_s^{(2)})^2 r_h^\frac{4}{3} \mu^\frac{2}{3}$. 
\item For $k =3$, $g_{xx}(-u_s(1+\epsilon e^{i 0^-}))$ is real for $\epsilon \geq 0$, implying that $c^{(3)} = 6^\frac{1}{3}e^{-\frac{2\pi i}{3}} (u_s^{(3)})^2 r_h^\frac{4}{3} \mu^\frac{2}{3}$. 
\item For $k = 4$, $g_{xx}(u_s^\star(1+\epsilon e^{i 0^+}))$ is real for $\epsilon \geq 0$, implying that $c^{(4)} = 6^\frac{1}{3}e^{\frac{2\pi i}{3}} (u_s^{(4)})^2 r_h^\frac{4}{3} \mu^\frac{2}{3}$.  
\end{itemize}
Taking stock, we finally find that
\begin{equation}\label{RN_asymptotic_pred}
[u^n](B) \sim - 2^\frac{4}{3}\,3^\frac{1}{3}\,r_h^\frac{4}{3}\,\mu^\frac{2}{3} {{\frac{2}{3}}\choose{n}} |u_s|^{2-n} \left(1+ (-1)^n\right)\cos\left( \theta_s n - 2 \theta_s - \frac{\pi}{3} \right).     
\end{equation}
The asymptotic prediction \eqref{RN_asymptotic_pred} is consistent with the fact that $B$ is real and even. For this reason, in the following we work with $k \equiv \frac{n}{2}$, $b_k \equiv [u^{2k}][B]$. The $b_k$ coefficients are determined by a numerical computation. 

\begin{figure}[h!]
\begin{center}
\includegraphics[width=\textwidth]{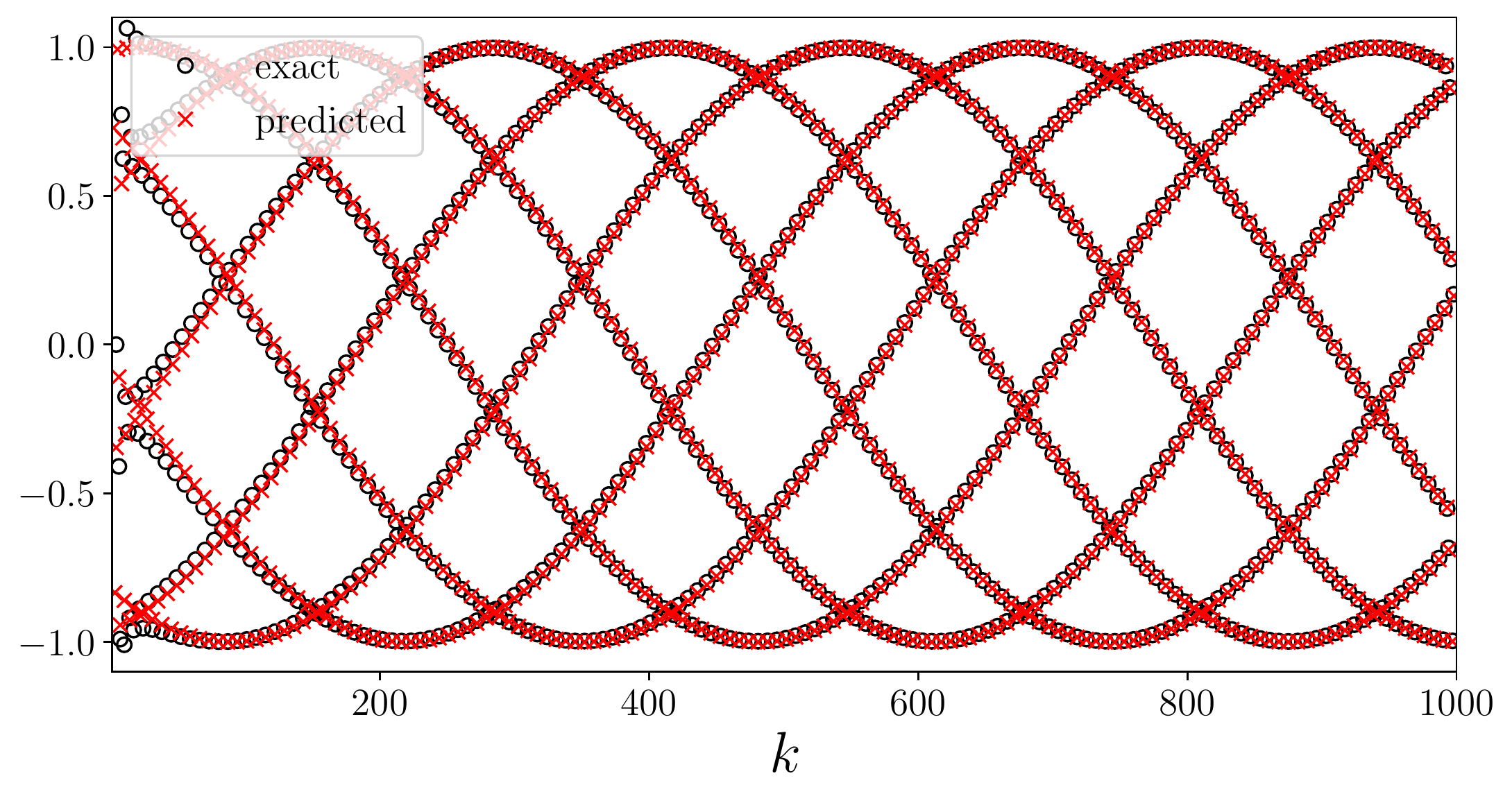}
\caption{For RNAdS$_5$ black branes, the large-order behaviour of $b_k = [u^{2k}](u^2 g_{xx})$ is governed by the black brane interior. In the plot, open black circles correspond to the ratio \eqref{RN_ratio} between the exact $b_k$ and the amplitude of the asymptotic prediction \eqref{RN_asymptotic_pred}. As discussed in detail in the text, at large $k$ this ratio has to tend to $\cos\left( 2\theta_s k - 2 \theta_s - \frac{\pi}{3} \right)$ (red crosses). The plot clearly shows that both quantities are in very good agreement at sufficiently large $k$. $|u_s|$ and $\theta_s$ have been determined by a numerical solution of the Einstein equations in the complex $u$-plane. The parameter values we have considered are $r_h =1.5$ and $\mu = \sqrt{3}$, resulting in $|u_s|\approx0.730$ and $\theta_s \approx 0.901$.}
\label{fig:Darboux}
\end{center}
\end{figure}

In figure \ref{fig:Darboux}, we plot
\begin{equation}\label{RN_ratio}
\frac{b_k}{-2^\frac{7}{3}\,3^\frac{1}{3}\,r_h^\frac{4}{3}\,\mu^\frac{2}{3} {{\frac{2}{3}}\choose{2k}} |u_s|^{2-2k}}    
\end{equation}
and compare it with asymptotic prediction $\cos\left(2\theta_s k - 2 \theta_s - \frac{\pi}{3} \right)$. $|u_s|$ and $\theta_s$ are obtained from a numerical solution of the Einstein equations in the complex $u$-plane. As the figure illustrates, for sufficiently large $k$ we get an excellent agreement between the exact result and the asymptotic prediction \eqref{RN_asymptotic_pred}. This demonstrates that the leading large-order behaviour of the Fefferman-Graham expansion is completely determined by the near-singularity behaviour of the metric in the complex $u$-plane. In particular, note that the ratio \eqref{RN_ratio} at large $k$ provides direct access to $\theta_s$, which corresponds to the proper time taken by a static observer to fall from the outer to the inner horizon. 

\section{Conformal maps and convergent expansions for black holes}
\label{sec:conformal_map}
\begin{figure}[h!]
\begin{center}
\includegraphics[width=\textwidth]{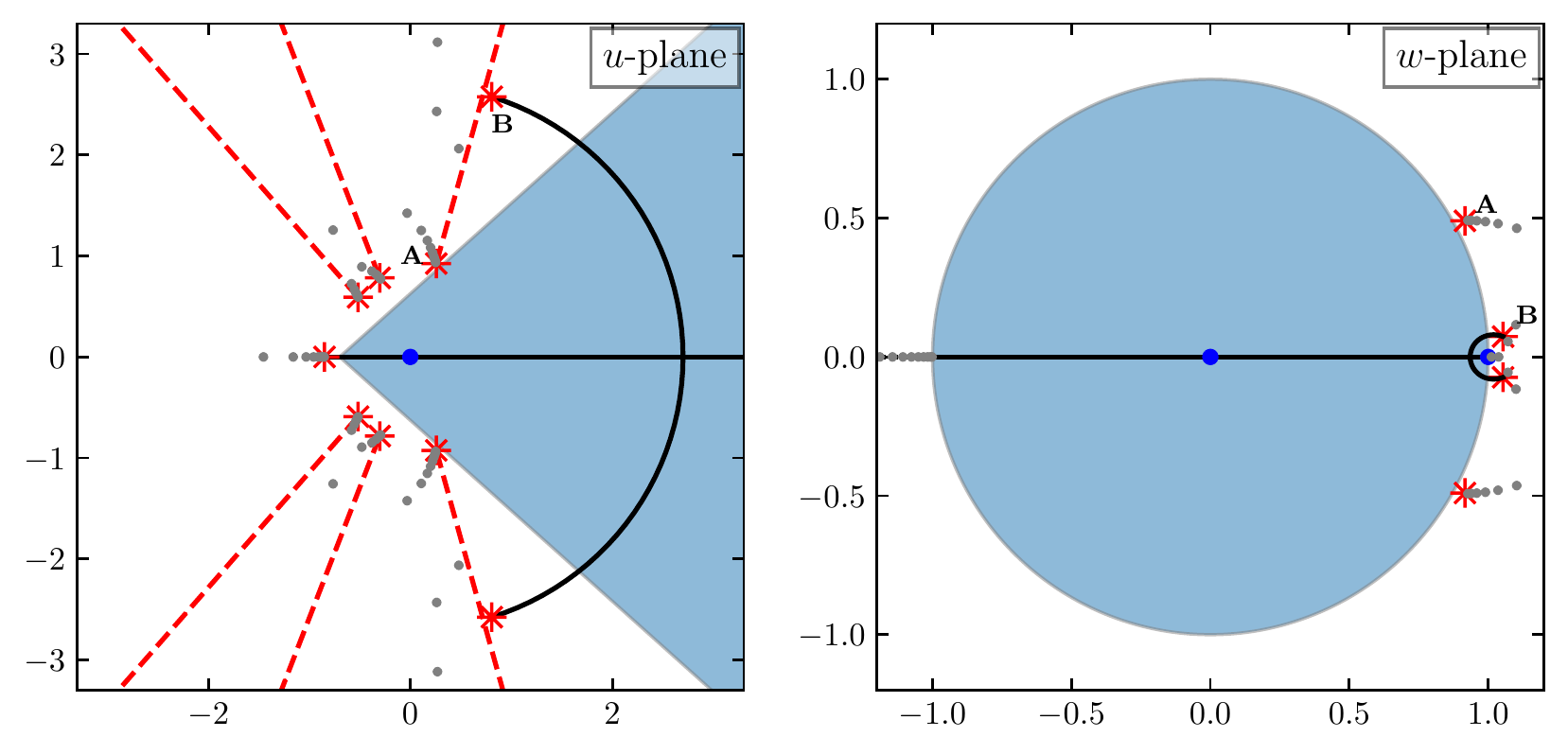}
\caption{Conformal map from Fefferman-Graham radial coordinate $u$ to a new radial coordinate $w$ defined by \eqref{wedgemap}. The wedge with opening angle $2\alpha$ and tip at $u=-\beta$ (blue region in the $u$-plane) is conformally mapped to the unit disk (blue region in the $w$-plane) with the conformal boundary at the origin. As such, a near-boundary expansion in the radial coordinate $w$ will converge up to and including a portion of the interior of the black hole, as well as the second asymptotic region. For illustrative purposes we have also plotted the rays, arcs and singularities for the holographic superconductor from section \ref{sec:HHH} on top of these regions with conformal map parameters $\alpha = \pi/4.3$ and $\alpha = 0.7$, and singularities \textbf{A}, \textbf{B} identified in both plots. The grey dots on the left are poles of the Pad\'e approximant of the $u$-expansion of $\psi$ (Fefferman-Graham), while the grey dots on the right are the same but for the $w$-expansion of $\psi$, further illustrating the absence of singularities inside the disk. There is a branch point at $w=-1$.}
\label{fig:conformalmap}
\end{center}
\end{figure}
Consider the conformal map $u\mapsto w(u)$, where
\be
w(u) = \frac{(\beta^{-1} u+1)^\frac{\pi}{2\alpha}-1}{(\beta^{-1} u+1)^\frac{\pi}{2\alpha}+1},\label{wedgemap}
\ee
with parameters $\alpha, \beta >0$. It maps a wedge in the $u$-plane whose tip is at $u=-\beta$ and opening angle $2\alpha$ to the unit disk in the $w$-plane, with the origin of the disk $w=0$ identified with $u=0$, and $u\to +\infty$ with $w=1$. We have therefore constructed a disk in the $w$-plane, inside which the small $w$ expansion converges and includes the second asymptotic region as well as a portion of the interior of the black hole. This map is illustrated in figure \ref{fig:conformalmap}. 
Since the $w$-expansion converges for at least some classes of black hole spacetimes it can be exploited as an alternative method to solve the Einstein equations. In practice, begin in radial gauge with radial coordinate $w$, and construct a near boundary expansion for small $w$. This converges in the unit $w$-disk provided the chosen $\alpha, \beta$ are not too large. Use this perturbative solution to impose the existence of a regular Killing horizon somewhere inside this disk by adjusting near-boundary data. The approximate solution in the Fefferman-Graham gauge can then be obtained by applying \eqref{wedgemap} exactly (that is, without expanding in $u$). We have verified that this procedure works for the model of section \ref{sec:HHH} by explicitly constructing the broken phase holographic superconductor with this method. Whether the $w$-expansion converges fast enough to be a practical approach compared to other techniques, remains to be seen.\footnote{Other techniques include using Pad\'e approximants of the Fefferman-Graham expansion, and carrying out analytic continuation by using Taylor series expansions at several points along the real $u$-axis and matching them together where the associated disks overlap.} It would also be interesting to understand whether this approach can be used to solve the Einstein equation PDEs for higher cohomogeneity examples. 

\bibliographystyle{ytphys}
\bibliography{fg}

\end{document}